\documentstyle[aps,prb]{revtex}
\begin{document}
\draft
\title
{Efficient ab-initio method for the calculation of frequency dependent
second order optical response in semiconductors}
\author{Sergey N.~Rashkeev\cite{a}, Walter R. L. Lambrecht and
Benjamin Segall}
\address{Department of Physics, Case Western Reserve University, Cleveland,
OH 44106-7079}
\date{\today}
\maketitle

\begin{abstract}
The development of a method for calculating 
the frequency dependent second harmonic
generation coefficient of insulators and semiconductors based 
on the self-consistent linearized muffin-tin orbitals (LMTO) 
band structure method is reported.
The calculations are at the independent particle level and are based
on the formulation introduced by Aversa and Sipe (1995). 
The terms are re-arranged in such a way as to exhibit explicitly
all required symmetries including the Kleinman symmetry 
in the static limit.
Computational details and convergence tests are presented. 
The calculated frequency dependent $\chi^{(2)}(-2\omega,\omega,\omega)$ 
for the zinc-blende
materials GaAs, GaP and wurtzite GaN and AlN are found to be in
excellent agreement with that obtained by other first-principles
calculations when corrections to the local density approximation (LDA)
are implemented in the same manner,
namely, using the ``scissors'' approach.
Similar agreement is found for the static values of $\chi^{(2)}$
for zincblende GaN, AlN, BN and SiC. 
The strict validity of the usual ``scissors'' operator implementation 
is, however, questioned. We show that better agreement 
with experiment is obtained when the corrections to the low
lying conduction bands are applied
at the level of the Hamiltonian, which guarantees
that eigenvectors are consistent with the eigenvalues. 
New results are presented for the frequency dependent 
$\chi^{(2)}(-2\omega,\omega,\omega)$
for 3C-SiC.
The approach is found to be very efficient and flexible which indicates
that it will be useful for a wide variety of material systems including
those with many atoms in the unit cell.
\end{abstract}

\pacs{PACS numbers: 42.65.Ky, 78.20.Bh}

\section{Introduction} \label{sec:int}
The calculation of non-linear optical susceptibilities 
of crystalline materials from 
first principles is an important but difficult task.
The increasing activity and realization of the enormous 
potential for technological applications 
in the area of non-linear optics (NLO), ranging from optical 
communications and computing to improved all solid state laser systems,
has recently heightened interest in this problem.
\cite{Levine1,Levine2,Levine3,DalCorso,Moss,GMS91,Ghah,Moss1,Huang1,Huang2,%
Sipe,Aversa,Hughes,Hughes1,Chen1,Chen2}
The materials requirements for 
certain NLO devices  are stringent and require a search for 
new materials and/or optimization of their NLO properties, for example by 
artifical structuring of materials such as superlattices and heterostructures.
Although the general principles of NLO response are 
understood, a little study of the literature reveals  that the 
uncertainties and the level of agreement between theory 
and experiment are still far worse than for linear optical response.
Hence, there is a great need to develop reliable quantitatively 
predictive methods to determine NLO susceptibitilies. This paper presents
a step in that direction. 

NLO also has a great potential as a characterization technique 
for materials,  because of its sensitivity to symmetry,
which stems from the fact that it is described by higher than rank 2 tensors.
As is well known for example, $\chi^{(2)}$ is zero for systems 
with inversion symmetry and thus any $\chi^{(2)}$ measured on such systems
must come from regions where the symmetry is broken such as surfaces
or interfaces. In order to extend the use of NLO spectroscopic 
characterization of interfaces of materials possessing 
an intrinsic $\chi^{(2)}$ in bulk, or to extract the 
maximal amount of information from such measurements, a more 
quantitative theoretical analysis will be  required. 
Requirements for such analysis are that the calculation of 
$\chi^{(2)}(-2\omega,\omega,\omega)$
must be feasible for models with large numbers of atoms used to 
represent bulk and surface or interface 
regions and must be physically transparent 
so that information on electronic states can be extracted
from it. The method presented here has been developed with these 
requirements in mind. 

Before going on to the more challenging studies of interface systems, 
it is necessary to establish the reliability of the methods for simple
bulk crystalline systems and establish what level of theory is required
by an iterative process between theory and experiment until a 
sufficient level of agreement is reached to extract useful information.
On the experimental side, some of the uncertainties stem from self-focusing,
imperfect phase matching, finite wave mixing, 
effects due to sample size and surface quality. In addition, special
equipment is required for the higher frequency UV response, 
the study of which is probably most readily done
using frequency doubling or tripling.
This explains perhaps to some extent why there are 
relatively few experimental 
measurements of the frequency dependent 
$\chi^{(2)}(-2\omega,\omega,\omega)$ extending 
into the absorbing region.
On the theoretical side, as in linear optics, one should in general 
be concerned with many-body effects, excitonic enhancements,
local field corrections, etc. Excitonic effects have been 
emphasized in the literature on NLO response in 
superlattices \cite{Agran,Tsang,Voon}  but are neglected
in the remainder of this paper because we find it necessary to first
establish the basic one electron picture firmly before going to these more
advanced topics. Local field corrections are briefly addressed below.
However, until recently, 
even at the most basic level of the independent particle
approximation, the formalism was by no means standard or universally 
accepted. The present paper will hence be mainly concerned with the theory 
at the independent particle level.

When comparing with linear response calculations, 
the novice in this area may  at first be somewhat bewildered by the 
many different formulations of the problem and the lack of a well-established
set of equations. A second barrier results from the complexity 
of the equations and the difficulty of transforming 
from one formulation to another or establish 
their equivalence or conditions  of applicability.
Although we do not intend at all to give a complete
overview here of previous activity in this field, we give a brief 
discussion in the following few paragraphs to allow us to describe
the context and limitations of our present approach.

From, the outset, we divide the methods in two large groups: the first
being restricted to the static limit  of $\chi^{(2)}$, or, at best,
with the non-absorbing frequency region \cite{Levine3}; 
and the second with the 
full frequency dependent method. The link between the two was 
for some time plagued by formal difficulties with apparently diverging
terms as discussed below. Therefore, early on,  separate   
semi-empirical methods were developed for the static limit. Most notably 
the bond orbital method used 
by B. F. Levine \cite{BLevine} was rather successful
at predicting static $\chi^{(2)}$ values for a wide class of materials.
Nevertheless it is limited in its scope and is outside the realm of 
first-principles methods envisioned here.
In recent years, the theory of the static limit of
the response functions was developed to a rather sophisticated level
in the framework of density functional theory. In particular, 
Levine  and Allan developed the method including self-consistently 
the local field effects.\cite{Levine1,Levine2,Levine3} 
Their formulation is  explicitly divergence free
by application of various sum rules and makes extensive use of rearranging
terms for maximal computational efficiency. These rearrangements, however
rely on remaining outside the absorbing frequency range 
in some instances and hence limit the applicability range of the method. 
Another disadvantage is that the  final formulas are quite lengthy
and thereby not easily related back to the underlying electronic states. 
In fact, they are so lengthy that their derivation and 
computational implementation made use of symbolic mathematical software. 
Very recently, Dal Corso et al. \cite{DalCorso} 
formulated the problem of 
calculating $\chi^{(2)}$ in terms of third order
derivatives of the total energy. By using the ``2n+1'' theorem 
\cite{Gonze89} they
were able to simplify the calculation of the static $\chi^{(2)}$ considerably
because only first order corrected wave functions are required. 
The latter are obtainable through a Sternheimer 
approach.\cite{Sternheimer,Baroni} 
Their formulation was extended to 
frequency dependent $\chi^{(2)}$ in the context of 
time dependent density functional theory, but, in practice, was again 
restricted to the non-absorbing regime. It by-passes 
some problems with the definition of the position operator for 
a periodic system by going to a Wannier function representation,
in a manner closely related to the Berry-phase theories of 
the macroscopic polarization.\cite{Vanderbilt,Resta}
The final formulas could of course 
be re-expressed in terms of Bloch functions.
Both Levine and Allan's and Dal Corso et al.'s methods were implemented
using norm-conserving plane wave methods and successfully applied to 
a variety of semiconductors.
Although these techniques are perhaps the most
advanced because of their inclusion of 
local field corrections, their extension to full frequency dependence
is not straightforward.  Since local field effects were found to
be generally only of the order of 10 \% for conventional semiconductors
we will neglect them in what follows. 
Although computationally convenient,  the use of the Sternheimer
approach instead of the usual perturbation theory blurs the link between 
optical response and interband transitions. Therefore, the interpretation of
optical response functions in terms of the underlying electronic band
structure, which is one of our goals, is encumbered. 

We next turn to the methods dealing with the full frequency dependence. 
After a few of the early attempts to apply perturbation 
theoretical techniques applicable to atoms and molecules 
to periodic solids, the earliest of which to our knowledge is 
the work of Butcher and McLean \cite{Butcher}, several 
problems became apparent. The equations appeared to be plagued by
divergences in the static limit. Aspnes \cite{Aspnes}
was the first to 
show that these were only apparent but his proof was limited 
to cubic crystals. For quite some time, this problem represented a 
serious setback to the full development of these methods 
in the context of band structure methods. 
The next important step was made by the work of Ghahramani, Moss and Sipe
\cite{GMS91} who gave a more general proof of the 
disappearance of the divergence
by using a new sum rule. Soon afterwards, Sipe and Ghahramani \cite{Sipe}
presented a more general approach for avoiding the divergences 
from the start by a careful
treatment and systematic separation of interband and 
intraband motion. Finally, Aversa and Sipe 
\cite{Aversa} showed that this could most easily be accomplished using 
a so-called length-gauge formulation. 
  
In this connection, perturbation theoretic approaches are most
frequently set up taking 
the interaction between
the long-wavelength electromagnetic field and the solid to be either  
($-(e/mc){\bf p}\cdot{\bf A}$), in which 
${\bf p}$ is the momentum operator and  ${\bf A}$ the vector
potential of the field,  or $e{\bf r}\cdot{\bf E}$, in which
${\bf r}$ is the position operator and ${\bf E}$ the electric field.
The former is often referred to as the velocity or momentum gauge
and the latter as the length-gauge formulation.
Of course, in the end ${\bf E}=-c^{-1}\dot {\bf A}(t)$ links the two 
approaches, which are equivalent through a unitary transformation 
as discussed explicitly by Aspnes \cite{Aspnes} and 
Aversa and Sipe.\cite{Aversa}
As explained in the 
papers by Sipe et al. \cite{Sipe,Aversa} each formulation has its
own advantages and disadvantages. The use of the position operator
has the problem that its matrix elements between Bloch functions
in a periodic solid are not defined uniquely.\cite{Blount} 
On the other hand, the use of the velocity-gauge formulation leads to 
additional frequency factors in the denominators and hence to the 
apparent divergences in the static limit, which must then be eliminated 
in the end by the use of suitable sum rules. 
A similar discussion appears in Ref.\onlinecite{Aspnes} although
some of the conclusions there are limited to cubic materials.
In the present paper, we adopt what seems at present to be the
most transparent formulation, namely the one given by 
Aversa and Sipe.\cite{Aversa} 
We were able to further re-arrange
the terms so as to explicitly display the so-called Kleinman 
symmetry \cite{Kleinman} 
in the limit $\omega\rightarrow 0$ \cite{note}. 
Equations in the suitably modified form are given in Sec. \ref{sec:form}.

The derivations by the Toronto group \cite{Sipe,Aversa} are given 
in the framework of density matrix equations of motion. We verified
that at least the pure interband terms for 
$\chi^{(2)}(-2\omega,\omega,\omega)$ 
can also be obtained using the Matsubara 
Green's function approach.\cite{Sergeyunpub} Nevertheless, we recognize
the advantages of the density matrix approach in obtaining a 
straightforward  hierarchical approach for higher response functions
such as $\chi^{(3)}$ which we intend to address in future work and 
for systematically including all intraband motion related 
terms. 

As far as practical band structure 
implementations are concerned, we note that until 
recently, most  of the calculations 
were performed using semi-empirical band structure methods.
The first attempts to calculate the static limit of the second order
nonlinear susceptibility $\chi^{(2)}(0,0,0)$, or $\chi^{(2)}$ for short, 
for semiconductors with zincblende structure were made under
the assumption that the momentum
matrix elements are constants in the Brillouin zone \cite{Aspnes}
and did not employ full band structures but only ${\bf k}\cdot{\bf p}$ 
expansions in the neighborhood of certain high symmetry points.
After this paper there were several attempts to obtain the second
and third order response functions from band structure calculations
involving a variety of band structure approaches
\cite{Fong,Moss,GMS91,Ghah,Moss1,Huang1,Huang2}, ranging from semi-empirical
pseudopotentials to tight-binding and  
self-consistent linear combination of atomic orbitals methods.
However, it was typically assumed 
that only the so-called virtual electron terms, $vcc^\prime$, involving 
one valence band ($v$) and two conduction bands ($c$ and $c^\prime$)
contribute 
appreciably. Also, the early work included only strictly three band terms, 
i.e. $c\neq c^\prime$, whereas the new formulations include some two-band
terms involving the quantity $\Delta^a_{nm}=(p^a_{nn}-p^a_{mm})/m$.
(Here and in the following, 
the superscripts $a$, $b$ and $c$ denote the Cartesian coordinates.)
It can be shown that these terms vanish  
for cubic materials, or, for example, for 
the $zzz$ component in hexagonal materials by time-reversal symmetry,
but they do not vanish in the general case. An important limitation of the 
semi-empirical band structure implementations is that it is not clear 
how well these approaches describe the wave functions and hence the matrix 
elements. It is rather difficult, in retrospect to disentangle the 
effects of the different expressions employed from the use of the 
necessarily imperfect matrix elements or other numerical limitations
of the calculations. Uncertainties about the magnitude of 
local-field effects, treatments of the gap underestimates typical of LDA,
and the uncertainties of the measurements themselves
add to the confusion. Only recently were the well founded 
and generally applicable formulations implemented in terms of 
an ab-initio local density functional theory based band structure method.
Hughes and Sipe \cite{Hughes,Hughes1} implemented the Aversa and Sipe 
approach using the full-potential linearized augmented plane-wave 
(FLAPW) method.\cite{OKA,flapw}
Using this formalism, the first accurate calculations 
of the full frequency dependent second order optical responses -
the second harmonic generation (SHG) coefficient and the 
linear electro optic (LEO) susceptibility -
were carried out for the zincblende crystals GaAs and GaP.\cite{Hughes} 
Subsequently, similar calculations of
GaN and AlN which have the wurtzite structure were performed.\cite{Hughes1}

In the work reported here, we chose to use the 
linearized muffin-tin orbital method (LMTO) \cite{OKA,lmto}
in the atomic sphere approximation (ASA).
We have found this method to be quite adequate for analyzing linear
optical response functions.\cite{Uspens,Alouani,uvsic,uvgan}  
It has the advantage of being computationally more efficient 
than the FLAPW method. This will allow us in future work to apply the 
method to more challenging systems with many atoms per unit cell
without compromising other convergence
criteria such as summations over intermediate states, Brillouin zone 
integrations and so on. Furthermore, it offers potentially 
more physical transparency in the sense that atomic orbital 
or angular momentum type decompositions could be
developed. Although the present implementation uses 
the ASA, this restriction could be removed
at a later stage without drastically changing the codes.  Finally, although
this is outside the scope of the present paper, 
we note that local field 
effects in linear response functions 
and approaches going beyond LDA have recently been implemented 
in a particularly efficient manner within an LMTO 
framework.\cite{Aryasetiawan}

Since this paper is the first in which we present results obtained with 
this approach, its primary aim is to demonstrate the 
accuracy and overall correctness of our computational method.
We discuss some of the associated technical problems, such as  the 
computation of the matrix elements and the inevitable convergence 
issues related  to various cut-offs associated with any numerical method,
such as the Brillouin zone integrations, the sums over intermediate
states, the angular momentum cut-offs in Sec. \ref{sec:comp}. 
Most of the convergence tests are presented for cubic SiC, because we 
have ample experience with this material from other studies
and because it is an interesting case to illustrate some aspects of the 
method such as the relative importance of the various terms 
in the equations because of the relatively simple two peaked 
shape of the linear response function. Also  its
interpretation in terms of interband transitions
is well-understood.\cite{uvsic} Our results for 
$\chi^{(2)}(-2\omega,\omega,\omega)$ for 3C-SiC are 
to our knowledge the first for this
material. However, we can compare  our results 
to Chen et al.'s work \cite{Chen1} in the static limit.
Our choice of the other 
materials for which results are presented is motivated
by the availability of results obtained by other research groups. 
In particular, we will present results 
for wurtzite and zincblende GaN because, 
we can compare both to the results obtained
by Chen et al.'s \cite{Chen2} approach in the static limit and with  
Hughes et al.'s results \cite{Hughes1} for the  frequency dependent case.
Accurate experimental results are also available for the static 
values \cite{Miragl} and the use of a hexagonal material demonstrates the 
applicability of our method to arbitrary crystal structures. 
GaAs and GaP were chosen because the availability of some frequency dependent 
experimental results\cite{Parsons,Bethune,Chang} 
and the prior treatments by various groups, including 
the work by Hughes and Sipe, to which our approach is most closely related.
These results and discussions of them are presented in Sec. \ref{sec:res}.
  
A separate issue to be discussed in connection with the methodology
is the use of the local density approximation for the calculation 
of the one electron band structure from which the 
response functions are calculated. 
This will be dealt with in Sec. \ref{sec:complda}. 
We will investigate the use
of the commonly used ``scissors operator'' correction \cite{LevAllan,Hughes} 
and a slightly different approach  in terms of a corrected Hamiltonian 
which we will show in Sec.\ref{sec:res} to have some advantages.

\section{Formalism for Second-Order Response Functions} \label{sec:form}

In this section, we provide a concise description of equations used
in this work
to calculate the second order response functions. These
equations were originally obtained by Sipe and Ghahramani,\cite{Sipe}
and later derived in a simpler manner by Aversa and Sipe,\cite{Aversa}
using the  above discussed length-gauge formalism.  
As mentioned in the 
introduction, we rearrange the equations slightly so as to make their 
symmetries under permutation of the indices more apparent.  
For the SHG in a system with filled bands 
(an insulator or clean semiconductor at zero temperature), one 
thus obtains,

\begin{equation}
\chi^{abc}(-2\omega,\omega,\omega)=
\chi_{e}^{abc}(-2\omega,\omega,\omega)+\chi_{i}^{abc}(-2\omega,\omega,\omega),
\label{eq:SHG_tot}
\end{equation}
where 

\begin{equation}
\chi^{abc}_e(-2\omega,\omega,\omega)=
\frac{e^3}{\hbar^2\Omega}
\sum_{nml,{\bf k}} \frac { r^a_{nm} \{r^b_{ml} r^c_{ln} \} }
{(\omega_{ln}-\omega_{ml}) }
\left[ \frac{2f_{nm}}{\omega_{mn}-2\omega} + \frac{f_{ln}}{\omega_{ln}-\omega}
+\frac{f_{ml}}{\omega_{ml}-\omega} \right]
\label{eq:SHG_e}
\end{equation}
is the contribution of the purely interband processes 
(all the three band indices $n,m,l$ being  different
since $r^a_{nm}$ are defined to be zero unless $n\neq m$) with 
$\{r^b_{ml}r^c_{ln}\}=(1/2)(r^b_{ml}r^c_{ln}+r^c_{ml}r^b_{ln})$, and

\begin{eqnarray}
\chi^{abc}_i(-2\omega,\omega,\omega)=\frac{i}{2} \frac{e^3}{\hbar^2\Omega}
\sum_{nm,{\bf k}} f_{nm} [
\frac{2}{\omega_{mn}(\omega_{mn}-2\omega)}
r_{nm}^{a}(r_{nm;c}^b+r_{mn;b}^c) 
\nonumber\\
+\frac{1}{\omega_{mn}(\omega_{mn}-\omega)}
(r_{nm;c}^{a}r_{mn}^b+r_{nm;b}^ar_{mn}^c)
\nonumber\\
+\frac{1}{\omega_{mn}^{2}} \left( \frac{1}{\omega_{mn}-\omega} -
\frac{4}{\omega_{mn}-2\omega} \right)
r_{nm}^a (r_{mn}^{b}\Delta_{mn}^c+r_{mn}^c\Delta_{mn}^b)
   \nonumber \\
-\frac{1}{2 \omega_{mn}(\omega_{mn}-\omega)}
(r_{nm;a}^br_{mn}^c+r_{nm;a}^cr_{mn}^b)  ],
\label{eq:SHG_i}
\end{eqnarray}
is the contribution of the mixed interband and intraband 
processes. The latter include
the modulation of the linear response by the intraband motion 
(in the first 3 terms, which correspond to $\eta^{abc}$ in 
Ref. \onlinecite{Hughes} or $\chi^{(2)}_\chi$ in Ref. \onlinecite{Aversa})
as well as
the modification of the intraband motion by the interband polarization
processes (last term corresponding to   $i\sigma^{abc}/2\omega$ 
in Ref. \onlinecite{Hughes} or  $\chi^{(2)}_\sigma$ 
in Ref. \onlinecite{Aversa}).
Here ${\bf r}$ 
is the position operator, $\hbar\omega_{nm}=\hbar\omega_n-\hbar\omega_m$,
is the energy difference for the bands $m$ and $n$, 
$f_{nm}=f_n-f_m$ is the difference of the Fermi distribution functions,
sub or superscripts $a$, $b$ and $c$ are Cartesian
indices, $\Omega$ is the unit cell volume and the frequency 
$\omega$ is to be understood as $\omega+i\delta$ for which the limit
$\delta\rightarrow+0$ is to be taken in the end.
All the quantities in the summation except $\omega$ depend on ${\bf k}$,
but this index is suppressed.
Finally, $r_{nm;a}^{b}$ is the so-called 
generalized derivative of the coordinate operator
in ${\bf k}$- space, 

\begin{equation}
(r_{nm}^{b})_{;{\bf k}^a}=\frac{r_{nm}^a \Delta_{mn}^b+r_{nm}^b \Delta_{mn}^a}
{\omega_{nm}} +\frac{i}{\omega_{nm}} 
\sum_l (\omega_{lm}r_{nl}^a r_{lm}^b - \omega_{nl} r_{nl}^b r_{lm}^a ) ,
\label{eq:gen_der}
\end{equation}
where $\Delta_{nm}^a = (p_{nn}^a - p_{mm}^a)/m $ is the difference between
the electronic velocities at the bands $n$ and $m$.
These equations have already been reduced to a form displaying single 
$\omega$ or $2\omega$ resonances, rather than products of two such energy
denominators. This facilitates the separate calculation of
the imaginary part, and, subsequently, the real part by a Kramers-Kronig
transformation. The equations can be further reduced by the substitution of 
Eq.(\ref{eq:gen_der}), which then leads to purely two-band 
terms involving the $\Delta^a_{mn}$ and additional three-band terms,
as explicitly written out in the Appendix B of Ref. \onlinecite{Hughes}.
We left the equations in this intermediately transformed state
(i.e., still involving some $r^a_{nm;b}$)  
and combined the two types of intraband terms because that 
made it easier to obtain a fully symmetric form in the static 
limit. After some algebra,
the Eqs. (\ref{eq:SHG_e}) and (\ref{eq:SHG_i})
can be simplified for the zero-frequency limit to give,

\begin{equation}
\chi^{abc}_{e}= 
\frac{e^3}{\hbar^2\Omega}
\sum_{nml,{\bf k}} \frac { r^a_{nm} \{r^b_{ml} r^c_{ln} \} }
{\omega_{nm}\omega_{ml}\omega_{ln} }
[\omega_{n} f_{ml}+\omega_{m} f_{ln}+ \omega_{l} f_{nm} ],
\label{eq:SHG_e0}
\end{equation}
and

\begin{equation}
\chi^{abc}_{i}= 
\frac{i}{4} \frac{e^3}{\hbar^2\Omega} 
\sum_{nm,{\bf k}} \frac{f_{nm}}{\omega_{mn}^2}
[r_{nm}^{a}(r_{mn;c}^b +r_{mn;b}^c) +
 r_{nm}^{b}(r_{mn;c}^a +r_{mn;a}^c) +
 r_{nm}^{c}(r_{mn;b}^a +r_{mn;a}^b) ].
\label{eq:SHG_i0}
\end{equation}

From these forms, 
it is obvious that both the 
$\chi^{abc}_e$ and $\chi^{abc}_i$ separately 
automatically satisfy the full permutation, or Kleinman 
symmetry.\cite{Kleinman} 
This  symmetry does not follow automatically 
from the point group
symmetry requirements for a tensor of a given rank in the crystal
and is only valid in the static limit. General arguments for its 
validity have been given elsewhere,\cite{Franken} although 
their validity as proof is not as clear as the present straightforward
derivation. 

The relative importance of the various terms will be discussed below in 
Sec. \ref{sec:res}. We anticipate that they are in fact of comparable
order of magnitude and often of opposite sign. Thus, extreme care 
is required in their computation and leaving out 
some of these terms may easily destroy results such as the Kleinman
symmetry which follows only from a delicate balance between these
terms. We note that the terms which involve only two bands, i.e.
involve the quantities $\Delta_{mn}$, are often relatively small.
In any case for cubic materials, their sum over the Brillouin 
zone vanishes identically. The 
terms labeled here as interband and intraband  are generally speaking 
of similar magnitude but of opposite sign as already discussed by 
Aspnes.\cite{Aspnes}

For nondegenerate bands $n$ and $m$ the matrix element
of the coordinate operator,
${\bf r}_{nm}$, is related to the momentum matrix element by
${\bf r}_{nm}={\bf p}_{nm}/im\omega_{nm}$, as follows from the equation of
motion. When the bands are degenerate, one can always choose the
wave functions $\psi_n$ and $\psi_m$ in such a way that a given 
component of ${\bf r}_{nm}$ or ${\bf p}_{nm}$ vanishes as noted in
Ref.\onlinecite{Aversa}. Thus, the degenerate bands do not 
contribute to  the response function. However, if the energy
difference between the two bands becomes small in some region 
of {\bf k}-space,  numerical problems can arise. 
We will return to this issue  in Sec.~\ref{sec:res}.

\section{Computational details, numerical accuracy and convergence tests} 
\label{sec:comp}
\subsection{General considerations}
In this section, we provide some further details on the computational 
approach and discuss some questions of numerical accuracy.
The general scheme  and set of approximations, or working assumptions,
have already been explained in Sec. \ref{sec:int}
and \ref{sec:form}.  To recapitulate, we work at the 
independent particle level with  the single particle spectrum
obtained within density functional theory in the local density 
approximation. Specifically, we use the Hedin-Lundqvuist 
parametrization of exhange and correlation here.\cite{HedinLund}  Some 
ad-hoc corrections  to the band structure are eventually applied to correct
for the well-known failure of the local density approximation 
in the LDA. Their justification and implementation will be discussed at 
the end of this section in subsection \ref{sec:complda}. The task at hand 
is then to evaluate the expressions given in Sec. \ref{sec:form}
with as input the band structure 
eigenvalues  and eigenfunctions of the one electron 
Schroedinger equation with the Kohn-Sham effective potential.  
More specifically, we need to discuss
the accuracy of the eigenvalues and the matrix elements and the 
convergence issues associated with the various summations, that is summations 
over band indices and Brillouin zone integrations.

First of all, as already mentioned, the calculations 
of the frequency dependent responses proceed (as is usual also 
for linear optics) 
by  first calculating the imaginary parts of Eqs.(\ref{eq:SHG_e}) and 
(\ref{eq:SHG_i}).  The real part is 
subsequently obtained by a Kramers-Kronig transformation. 
Each term in the imaginary part 
takes the form familiar from linear optics 
of a valence to conduction band transition but is now 
multiplied by a factor which in the three band terms 
contains a sum over intermediate states and consist 
of some combination of matrix elements divided by an 
additional energy difference denominator.

For reasons mentioned in the introduction, 
we chose the LMTO method for calculating the bands and the matrix elements.  
As most other modern band structure methods,
this  method is based on a variational approach using energy independent 
basis sets. As in all variational principle methods,
the error in the wave functions (and therefore, for the transition
matrix elements) are larger than the errors in the eigenvalues. Consequently,
the accuracy of the band structure alone can not be considered
to be an adequate criterium of the accuracy of the matrix elements and  
the optical properties derived from them.
A rather general discussion of this problem was given by Uspenski 
et al. \cite{Uspens}. They showed that the basic origin for the problem
is that the usual definition of the current density operator 
matrix elements which enter the response functions
as $\langle {\bf k}n|(\hbar\nabla/im)|{\bf k}n^\prime\rangle$
needs correction terms 
when the basis set is not complete. However, no practical method 
is available for computing these terms.  The 
practical recommendation of that paper is to estimate the 
magnitude of the required corrections by checking the fulfillment
of sum rules. This will be discussed further below in Sec. \ref{sec:methsum}.

\subsection{LMTO aspects} \label{sec:methlmto}

In the so-called linear methods, including in fact both the LMTO 
and FLAPW methods, \cite{OKA,lmto} the energy independent basis sets 
are constructed by expanding the partial waves centered
about atomic sites about a
chosen energy $E_\nu$ to first order in $(E-E_\nu)$. 
The errors  in the 
eigenvalues and eigenvectors increase respectively as
$(E_{{\bf k}\lambda} -E_{\nu})^4$ for
the energies and as $(E_{{\bf k}\lambda} -E_{\nu})^2$ for the wave
functions.  Typically, for self-consistent 
calculations $E_\nu$ is chosen in the center of the occupied
bands so as to give the best overall charge density. This linearization 
procedure is typically acceptable within a range of about 1 Ry below
and above the Fermi level. Because the matrix elements between the occupied
states and the conduction band states  rapidly decrease with increasing 
energy, this limitation to low lying states is of no great
concern in linear optics.  In the expressions of NLO, however, 
matrix elements between two conduction bands appear and therefore the 
higher states could possibly play a more important role in the intermediate
state summation. Therefore we must be concerned about their accuracy.
Recently, Aryasetiawan and Gunnarsson \cite{Ferdi} showed that this 
problem can be overcome by defining a multiple LMTO basis set which 
uses several linearization energies. They found this to be important for 
Ni when dealing with the energy loss spectrum in the region of 50-100 eV.
In the present applications to semiconductors, however, we found 
intermediate summations to converge rather rapidly, typically including 
bands only about 1-2 Ry up in the conduction band. Therefore, we did not find 
it necessary to implement this extension of the LMTO method at the present 
time.  It could be done at a future time when found to be necessary
for more demanding applications.

Another possible concern in connection with the LMTO method is our use of the 
ASA. For the semiconductors we typically find
that eigenvalues are in agreement with full-potential LMTO
calculations to a precision of 0.1 eV. The occasional exceptions occur only
at particular {\bf k}-points and therefore do not 
seriously affect the overall response functions which are obtained
from full Brillouin zone integrations.   

We now turn to the accuracy of the matrix 
elements. A first remark is that while in linear optics 
only the absolute values of the matrix elements appear and thus all 
contributions add with the same sign, this is not the case in NLO. 
Products of  three matrix elements appear 
and can have either sign. Large cancellations occur and thus 
more care is required. 
As already mentioned, the dipole (or position) matrix elements 
are avoided by rewriting all equations in the end 
in terms of momentum matrix elements whose treatment is better established.
A problem occurs in evaluating the matrix elements $r^a_{mn}$ 
as $p^a_{mn}/im\omega_{mn}$ when the bands $n$ and $m$ are nearly degenerate.
Consistent with the fact that the matrix elements can 
be made to vanish when the bands are exactly degenerate, 
we set $r^a_{nm}$=0 
when $\hbar\omega_{mn}\le\epsilon$ with the small arbitrary
cut-off $\epsilon$ chosen, e.g., to be 10$^{-3}$ Ry. We checked numerically
that our results are insensitive to the actual choice providing it is 
small. When we make it too small  the corresponding contributions 
can become too large and lead to unphysical spikes in the spectra.
The above choice is generally found to avoid such occurences. 

The momentum matrix elements  are calculated from the so-called 
one-center partial wave expansions of the LMTO method.  
As is well known, within the LMTO method, one can either expand
the total wave functions in multi-centered muffin-tin orbitals
\begin{equation}
\chi^{\bf k}_{RL}=\phi^{\bf k}_{RL}-
\sum_{R^\prime L^\prime} \dot\phi^{\bf k}_{R^\prime L^\prime}
 S_{R^\prime L^ \prime,RL}({\bf k}), \label{eq:lmto}
\end{equation}
in which $S_{R^\prime L^ \prime,RL}({\bf k})$ are the structure constants and 
$\phi^{\bf k}_{RL}$ and $\dot\phi^{\bf k}_{RL}$ are 
atom-centered partial waves 
(i.e. radial functions times spherical harmonics) at 
energy $E_\nu$ and their energy derivatives respectively,
or in a one-center expansion in terms of the 
$\phi^{\bf k}_{RL}(E_{n{\bf k}})=\phi^{\bf k}_{RL}+(E_{n{\bf k}}-E_\nu)
\dot\phi^{\bf k}_{RL}$.
Here, $R$ labels the atoms in the unit cell and a Bloch summation 
has already been carried out, 
and $L=\{l,m\}$ labels the spherical harmonic components. 
Within the so-called nearly orthogonal representation \cite{lmto} 
the same expansion coefficients $C^{n{\bf k}}_{RL}$ occur in both 
expansions of the Bloch eigenstates 
\begin{equation}
\psi_{n{\bf k}}=\sum_{RL} C^{n{\bf k}}_{RL}\phi^{\bf k}_{RL}(E_{n{\bf k}})=
\sum_{RL} C^{n{\bf k}}_{RL}\chi^{\bf k}_{RL}.
\end{equation}
The matrix elements between Bloch states thus reduce 
straightforwardly to internal products of the eigenvectors of the 
LMTO secular equation and 
matrix elements between the atom centered partial waves.
By expanding the momentum operator $i\hbar\nabla$ 
in its  spherical tensor components $i\hbar\nabla_m$ 
with $m=0,\pm1$, given by $\nabla_0=\partial/\partial z$, and
$\nabla_{\pm1}=\mp(1/\sqrt{2})(\partial/\partial x\pm i\partial/\partial y)$,
the latter matrix elements in turn 
reduce to radial matrix elements and Clebsch-Gordan coefficients.  

The concerns about the numerical accuracy arise from the need to truncate the 
above expansions in angular momentum and from the ASA in reducing 
all integrals over full space to a sum of integrals over space-filling 
spheres. One center expansions are known to require higher angular
momentum cut-offs than the multi-centered expansion. Typically,  
high accuracy linear optical applications require 
a $l_{max}=3$  for systems (including most semiconductors)
in which normal self-consistent 
calculations of the energies require only an 
$l_{max}=2$. While in linear 
optics this mainly affects absolute values and the 
correct spectral shapes can be obtained with even  smaller cut-off,
the inclusion of $f$-waves is imperative for NLO because
of the varying sign of different contributions. 
We note that this does not necessarily require one to increase the basis set 
to $f$-waves. In other words, the matrices required to be diagonalized 
do not have to increase in dimension. 
It is only the one-center expansion via 
structure constants, in other words, the $L^\prime$ summation in 
Eq.(\ref{eq:lmto}) that needs to be taken to a higher cut-off. The
procedure for obtaining the higher partial wave components of the 
partial wave expansion is known as ``blowing-up''.\cite{lmto}
An alternative way of including these higher components approximately 
and thereby insuring the continuity of the wave functions across
the boundaries of the atomic spheres 
while at the same time correcting for the geometric errors of the 
ASA (due to the overlap region of the spheres
and the neglect of regions of space outside all spheres) is provided
by the so-called combined correction technique.\cite{lmto}
Uspenski et al. \cite{Uspens}
have investigated how this affects the matrix elements 
in LMTO linear optics calculations. They found the corrections to be 
important only when a small $l_{max}=2$ is used and to be fairly small 
for the important low lying bands. Alouani et al. \cite{AlouaniKoch}
also showed that the combined  correction effects overall  
are small for the linear optical response functions.
We have not implemented these corrections 
for the matrix elements because we want to avoid the 
use of the plane wave expansion used in these earlier 
implementations of the  combined correction and because the 
use of a higher $l_{max}$, which is a simpler procedure, seems to 
be equally effective.  
We note that the combined correction to the overlap matrix 
is included, in a somewhat different form in setting up the 
secular matrix, and thereby improves the accuracy of the eigenvalues.
The corrections to the matrix elements can be implemented in a similar 
manner and this is planned for future work. 
The good agreement of our results with those of the FLAPW 
based calculations \cite{Hughes,Hughes1}, which 
handle the interstitial region more accurately, suggests that 
the above mentioned procedure does not represent
an important limitation. 

\subsection{Sum-rule tests}\label{sec:methsum}

As discussed by Uspenski et al.,\cite{Uspens} one way to estimate 
the overall accuracy of the matrix elements,  is to 
check the fullfillment of certain sum rules. 
Namely, the first partial derivatives
of the band energies in the {\bf k} space must give
the electron velocity,

\begin{equation}
\frac {\partial E_{\bf{k}\lambda}} {\partial \bf{k}_{\alpha}} =
\frac{\hbar}{m} \bf{p}^{\bf{k}}_{\lambda\lambda},
\label{eq:sum1}
\end{equation}
while the second derivative provides the inverse effective mass,

\begin{equation}
\hbar^2
\left( \frac{1}{m^{*}} \right)_{\alpha\beta}
=\frac {\partial^{2} E_{\bf{k}\lambda}}
{\partial \bf{k}_{\alpha} \partial \bf{k}_{\beta}}=
\frac{\hbar^2}{m} \left[ \delta_{\alpha\beta} + \frac{1}{m}
\sum_{\lambda' \neq \lambda}
\frac{\bf{p}^{\bf{k},\alpha}_{\lambda\lambda'}
      \bf{p}^{\bf{k},\beta}_{\lambda'\lambda}+
      \bf{p}^{\bf{k},\beta}_{\lambda\lambda'}
      \bf{p}^{\bf{k},\alpha}_{\lambda'\lambda}}
     {E_{\bf{k}\lambda}-E_{\bf{k}\lambda'}} \right] .
\label{eq:sum2}
\end{equation}
Inspection of the terms in Eqs.(\ref{eq:SHG_e}) and (\ref{eq:SHG_i}) 
reveals that very similar summations appear as those in 
Eq.(\ref{eq:sum2}), involving symmetrized products of two momentum 
matrix elements divided by some energy denominator and multiplied 
by another matrix element. Therefore, we feel that the second of these
sum rules provides a good test of the overall accuracy of the matrix 
elements. We already know from previous work \cite{Uspens} that 
including higher angular momenta (in practice up to $l_{max}=3$) 
and/or combined corrections improve the satisfaction of the first sum rule.   

We carried out some of these tests for various semiconductors 
in connection with independent work on effective masses.\cite{masgan,sicunpub}
As an illustrarive example, 
Table \ref{tab:mass} shows the calculated values of the effective masses 
for SiC at the $X$-point of the BZ along the $\Gamma-X$ direction for several
valence and conduction bands.  The masses were calculated 
directly from the band curvatures and by using the right
hand side of Eq.(\ref{eq:sum2}) up to various angular momentum cut-offs.  
The number of bands included was 10 but we found that generally 
only 3 to 4 terms contribute significantly.
The overall agreement between the masses calculated from the momentum
matrix elements and from the direct differentiation of the band energies
in the ${\bf k}$ space is satisfactory, the maximal deviation 
being about 0.04 electron mass units and typically deviation
being smaller than 10 \%. 
We hasten to add that here we only check internal consistency. 
Slightly different masses are obtained when FP corrections are included
and furthermore, there are errors to be expected from the LDA.
Our most important concern here was to ascertain the $l_{max}$ 
we expect to need for the calculation of NLO coefficients. 
Clearly, the convergence in terms of $l_{max}$ is better than
we had anticipated. There remain the more general limitations on the 
fullfilment of these sum rules, which are unavoidable  in any variational
finite basis set method, as discussed above.
The present tests  indicate that  these lead to 
an overall error  of the order of 10 \%.

\subsection{Further convergence tests}

We next study 
the convergence for the static values of 
$\chi^{(2)}$ for cubic SiC.  Table \ref{tab:conv} 
shows the LDA values of the SHG in the zero frequency
limit for different maximal value of the angular momentum
($l_{max}$) included in the LMTO one-center 
partial expansion of the Bloch eigenfunctions and for different numbers
of bands ($n_{max}$) included in the sums. We choose to carry out all 
sums over the same number of bands and $n_{max}$ refers to all bands,
including both valence and conduction. 
  
The conduction band width is about 0.8 Ry for 10 bands, 
1.7 Ry for 15 bands and increases to about 4.5 Ry for 30 bands. 
In spite of the fact that the LMTO method simply cannot give 
the correct results for the electronic bands at such high energies,
one can see that the values of $\chi^{(2)}$ are well-converged
for $n_{max}\ge15$ (remaining 
changes being smaller than 2.5 \%) and are even reasonable for $n_{max}=10$.
This indicates that the accuracy of the higher bands is not crucial
because the important contributions are from the lower bands. 
This results from the additional energy denominator prefactors which
cut-off the higher bands more rapidly than in linear optics or in 
the effective mass sum rule. We found that the individual terms 
converge somewhat more slowly: the cancellations between intra- and 
interband contributions help the convergence.

As for $l_{max}$ convergence, $f$-orbitals are clearly needed
while $g$-orbitals can be neglected.
It is not surprising that the calculations with $l_{max}$=2
gives an underestimate for the static limit of $\chi^{(2)}$.
In our earlier papers\cite{uvsic} on the linear response functions of 
SiC polytypes, we used $l_{max}=2$. However, this work was mainly 
aimed at identifying the prominent interband transitions and comparison 
to reflectivity measurements rather than at the absolute values. 
We recently found that with this $l_{max}$ the 
dielectric function is underestimated by 20--30\% \cite{pstatsol}.
The $f$-bands for SiC have very high energies, and the partial
contribution of the $f$-states in the density of states for the bands
around the semiconducting gap is relatively small.
However, the radial matrix elements between the $d$- and $f$-states
are large enough to create a noticeable $d-f$ contribution to the
matrix elements even in the gap region. 
In the case of the SHG, ignoring the $f$-states in the basis
orbital creates even larger differences.

\subsection{Brillouin zone integrations}\label{sec:methbz}

Finally, we discuss the Brillouin zone integration. For this purpose
we use the linear tetrahedron method. 
The integrations are converted into a weighted sum using the appropriate
Fermi cut-offs inside each microtetrahedron. The sums are 
reduced to an irreducible set of ${\bf k}$-points, rather than defining
a compact (but sometimes complicatedly shaped) irreducible 
portion of the Brillouin zone as discussed by Bl\"ochl et al.\cite{Blochl} 
The irreducible sets are obtained
by applying all symmetry operations of each 
${\bf k}$-point's applicable point group, thus generating its entire
star before integration.   
These leave the eigenvalues invariant and transform 
the (complex) matrix elements in a well-known
fashion. While time reversal symmetry still allows us to relate
quantities at ${\bf k}$ and $-{\bf k}$, their application is a little 
more subtle than in linear optics where only absolute values 
of the matrix elements appear. Here, we have ${\bf p}_{mn}(-{\bf k})
=-{\bf p}_{nm}({\bf k})=-{\bf p}^*_{mn}({\bf k})$. 
The symmetry aspects of our code were checked in various ways, for example,
by performing the calculation with and without certain symmetry operations
(integrating in the latter case explicitly over the full BZ),
and checking their consistency. We also calculated results for zincblende
treating it as a system with rhombohedral symmetry and found the required 
relations between the tensor components in rhombohedral axes 
and in cubic axes (to be discussed below) to be fulfilled exactly. 

Because of the additional
rapidly varying energy difference denominators  and the varying signs 
of the matrix element products,
a finer {\bf k}-point mesh is required than in linear optics.
We find that the difference between 
735 and 1456 irreducible {\bf k} points 
is about 5\% on average for the frequency dependent SHG
in crystal with zincblende structure. While the static 
limit can be obtained from the Kramers-Kronig transformation, it is more 
efficient to implement directly the simplified expressions of Eqs.
(\ref{eq:SHG_e0}) and (\ref{eq:SHG_i0}).
The consistency between  both approaches is found 
to be fulfilled numerically to better than 5 \%.
In the direct calculation of the static limit, 
we calculate analytically the contributions of each microtetrahedron
by linearizing  the energy bands in the denominators as well as the 
product of three momentum matrix elements in the numerator. 
This generally leads to integrands of the form $f({\bf k})/g({\bf k})$
with both $f$ and $g$ linear functions, which can easily be evaluated
analytically and expressed in terms of the 
corresponding quantities at the four corners of the tetrahedron. 
Such a scheme is superior to the usual averaging of the integrand
values in the corners of tetrahedron which becomes exactly equivalent to 
straigthforward sampling in the case of completely filled bands.
The need for an accurate  integration scheme involving a 
linear  interpolation of both the matrix elements and the eigenvalues
was also mentioned by Hughes and Sipe \cite{Hughes} who used a 
random sampling inside each microtetrahedron. Our 
semianalytic technique is believed to be more efficient.
A similar technique was described by Moss, Sipe and van Driel.\cite{Moss}
The formulas for the zero frequency limit
are less singular, and 300--500 {\bf k}-points
were found to be sufficient for obtaining  convergence to within 5\%.

\subsection{LDA errors} \label{sec:complda}

The use of the LDA 
is well-known to underestimate band gaps in semiconductors.
Since these energy band differences enter the calculations of response
functions in the denominators of the expressions, it is not 
surprising that the latter tend to be overestimated. 
This is well-known to lead to errors
of the order of 10-20 \% in linear response. 
The proplem is aggravated in higher 
order responses by the fact that denominators occur in higher powers.
It easily leads to factors of 2 or more in $\chi^{(2)}$. 
One might at first think that only an upwards 
shift of the conduction bands is needed. However, 
it has been argued that, at least, in the static limit, the response
functions are ground state properties and therefore should 
be based on the Kohn-Sham eigenvalues of density functional theory 
and not on the quasiparticle band structures. It is commonly 
accepted that the two differ not so much as a result of 
the local density approximation  but rather by the gap 
discontinuity\cite{Godby}  between the $N+1$, and $N$ particle 
systems. An alternative point of view, however, was presented by Gunnarsson
and Sch\"onhammer \cite{Gunnarsson}. They emphasize that the exact 
density functional may be extremely non-local. As a more practical
matter, Lambrecht and Segall
\cite{Lambrechtdiel}, and Levine and Allan \cite{LevAllan}
both found that corrected band structures provide improvements 
for static dielectric functions. Recently, it was shown
by Aulbur et al. \cite{Aulbur} that an alternative
justification for the use of a corrected band structure can 
be obtained from a recent extension of the density functional theory 
in which exchange correlation is considered to be an explicit functional 
of the polarization vector as well as the density \cite{Gonze95}
and thereby requires an extremely non-local term in insulators in 
the presence of an electric field. Thus, even for the static response 
a correction beyond LDA seems to be called for.

When we consider the frequency 
dependent response functions, it is clear that some correction to the 
band structure is necessary in order to have the resonances 
occur at the correct
energies. The simplest form of the correction is known as the 
``scissors'' approximation and consists of a 
shift of all the conduction bands by the gap correction $\Delta$
which in all practical applications
has been taken to be independent of {\bf k}. As was shown by 
Levine and Allan \cite{LevAllan}
and also, in perhaps a more transparent manner, by 
Hughes and Sipe \cite{Hughes}, this shift of the conduction bands 
- which represents a change in the Hamiltonian - requires
an associated renormalization of the momentum matrix elements

\begin{equation}
{\bf p}_{nm}\mapsto{\bf p}_{nm}\frac{\omega_{nm}+
(\Delta/\hbar)(\delta_{nc}-\delta_{mc})}{\omega_{nm}}, \label{pshift}
\end{equation}
where the ($\delta_{nc}-\delta_{mc}$)
factors restrict the correction to pairs of bands 
involving one valence and one conduction band state. Implicit in this
approach is the assumption that because LDA wave functions are close to 
the true quasiparticle wave functions (a result established, for example,
by Godby et al. \cite{Godby}
using quasiparticle calculations in the GW \cite{Hedin} approximation), 
the ${\bf r}_{mn}$ matrix elements do not change. 
We note that this assumption cannot be strictly correct.  
In fact, a consequence of the scissor approximation 
with a {\bf k}- independent $\Delta$ is that the 
effective masses remain unchanged. 
However, applying the above prescription for the momentum 
matrix elements to the well-known ${\bf k}\cdot{\bf p}$ expression for 
the inverse masses (see Eq.(\ref{eq:sum2})) 
increases the inverse mass tensor, or, decreases
the effective mass. 
It is also relevant to note that for most semiconductors, e.g.
GaAs, the LDA already underestimates the mass. 
(For a more complete discussion of the 
changes in the effective mass tensor due to quasiparticle corrections 
beyond LDA in the framework of the GW approximation, see Zhu et al.\cite{Zhu}).
Nevertheless, for optical response functions, the above 
prescription appears to be better than the alternative assumption that 
the ${\bf p}_{mn}$ matrix elements remain unchanged. This would 
strongly overcorrect the LDA error because for example in linear
response three energy denominators would change instead of only one.

We consider a slightly different, although also an admittedly ad-hoc 
correction which, however has the advantage of 
maintaining the consistency between eigenstates and Hamiltonian.  
This alternative approach is closely related to the one suggested
by Christensen.\cite{Christensen} It is essentially an empirical 
adjustment to the LMTO Hamiltonian chosen so as to reproduce the 
correct band structure. It  makes use of the fact that the lowest
conduction bands at $\Gamma$ and $X$  have a strong s-partial wave
character centered at the cation and 
(tetrahedral interstitital) empty sphere sites.
Therefore, we can simply introduce a shift of the corresponding
diagonal Hamiltonian matrix element. (Christensen achieves this 
by adding sharply peaked repulsive external potentials to the Kohn-Sham
effective one-electron potential in the corresponding spheres.
Such potentials primarily act on s-like states
because s-like states are the only ones with a non-vanishing 
probability at the origin. We find it slightly easier to 
make the corrections directly at the level of the 
Hamiltonian matrix, but the results of both approaches are the same.)
This scheme has the advantage that the
gap can be adjusted differently at different {\bf k}-points. On 
the other hand, it leaves the higher bands largely unchanged. In GaAs
and GaP, the 
important peaks in the linear response functions are dominated by the
transitions solely to the lowest conduction band. Therefore, this approach 
works very well, at least in the relevant energy range of these transitions.
It is here also shown to improve the agreement with 
experiment for the absolute magnitude of $\chi^{(2)}$.
It is not clear, however, how to generalize it so as to also shift up 
higher conduction states in cases where we think that is important. 
In fact, for SiC \cite{uvsic} and GaN,\cite{uvgan} 
we found in earlier work on linear response 
functions, that a rigid upward shift of all the conduction bands within
a range of about 10 eV above the minimum is required to obtain 
good agreement in peak positions with experimental reflectivity 
spectra. Hence, for SiC and GaN, we here use the scissor's approximation as 
suggested by Hughes and Sipe.\cite{Hughes} This also allows for a more
straightforward comparison to their results.

\section{Results} \label{sec:res}

\subsection{Static limit}
Table \ref{tab:GaAs} shows the calculated values of the SHG
for zincblende materials GaAs and GaP compared with other
theoretical calculations and available experimental data.
We show our LDA results as well as results of calculations with
gaps adjusted to the experimental values.
We used the ``scissors'' corrections to the LDA in the same
way as in Ref.\onlinecite{Hughes} as well as the shifts of cation and
empty sphere centered s-partial wave
diagonal matrix elements of the LMTO Hamiltonian,
which as discussed in Sec. \ref{sec:complda}
is equivalent to the addition of the short-range
potentials on the atomic sites.\cite{Christensen} 
As one could expect, the pure LDA values
of SHG are highly overestimated (especially for GaAs) 
because of the gross underestimate of the gap. 
The results for the ``scissors'' corrections are very close
to those of Hughes and Sipe \cite{Hughes} obtained by the self-
consistent FLAPW method. This verifies that the 
FLAPW and LMTO give essentially the same results for NLO
responses. The possible large errors in the band energies for
high lying bands which appear in any linear method
are not that important here because the main contributions
to SHG come from transitions to just a few 
low-lying conduction bands. 

In the present calculations we ignored the local field corrections.
However, they do not change the SHG in the zero frequency limit
by more than 10 \%
for these materials as follows from Ref.\onlinecite{Levine3} and 
as can also be seen in Table \ref{tab:GaAs}.
The more important correction is the gap adjustment. 
Table \ref{tab:GaAs} indicates that we get much better agreement with
the experimental results by applying the shifts in the Hamiltonian
than by applying a scissors correction to the final bands. 
The difference between the
approaches is that by explicitly
diagonalizing the shifted Hamiltonian
we obtain the correct
eigenvectors (and, therefore, the transition matrix elements)
of the ``shifted'' Hamiltonian, i.e., the eigenvectors 
and eigenvalues are consistent with each other.
This consistency is lost when one uses the ``scissors'' approach.
The LDA gap for GaAs obtained in ASA LMTO is very small (about 0.2 eV),
i.e., the ``scissors'' generates the gap rather than corrects it.
Therefore, we think that shifting the Hamiltonian matrix
elements is certainly a more suitable procedure than 
applying the ``scissors''
approach for this material.
For GaP the situation is much better from that point
of view because the LDA gap is much larger than for GaAs.
However, the ``scissors'' approach still gives an underestimation of
the experimental value of the SHG.
It is worth noting that Dal Corso et al. got a good agreement 
with the experimental results from their density functional
approach \cite{DalCorso} without any gap corrections.
However, their results depend strongly on whether or not they 
include the so-called non-linear core correction to exchange 
and correlation and their ``best results'', 
in terms of agreement with experiment,
are actually obtained without the correction.  There may be a 
compensation of errors occuring here.
In fact, the question of a gap correction does not enter 
in their approach because in their Sternheimer method,
the first-order corrected wavefunctions are obtained directly 
instead of in terms of the excited state wave functions of the zeroth 
order Hamiltonian as in the standard perturbation theoretical approach.
Thus, the conduction band does not enter their calculation. 
In view of the recent discussions of polarization dependent 
functionals,\cite{Gonze95}
one would think that a non-local correction to the 
exchange correlation is required.
We will return to discussion about different approaches to
the gap corrections in the next subsection.

The results of our calculations for some wide band gap semiconductors
with zincblende structure are shown in Table \ref{tab:zb}.
Here we used the ``scissors'' method of Ref.\onlinecite{Hughes} 
for simplicity. 
We see that in general the scissors significantly change the
absolute values of SHG. 
Our LDA values agree fairly well with
the LDA results obtained by the pseudopotential method.\cite{Chen1,Chen2}
The agreemement is particularly good for GaN taking into account that 
local field corrections are expected to reduce the value by about 10 \%.
Furthermore, this is in spite of the fact that
the pseudopotential calculations of Ref.\onlinecite{Chen2} did not
take into account the 3d bands of Ga while our LMTO calculations did.
It implies that these bands play a minor role in $\chi^{(2)}$.
For the other materials, applying an ad-hoc 10 \% reduction to our 
results due to local field 
corrections would worsen  the agreement.
In any case, Chen et al.'s \cite{Chen1,Chen2} and our calculations 
agree  on the very small value for AlN, the relatively small value
for BN, and the very close values for SiC and GaN.
The remaining discrepancies  are hard to pin down because of the 
hugely different computational procedure, possible use of different 
lattice constants, etc. Our calculations 
were performed at the experimental lattice constants.
For 3C-SiC we changed the sign of  the SHG 
coefficient given in Ref.\onlinecite{Chen1} 
because in the coordinate system 
used there the Si-C bond was oriented in the direction opposite
to that in our system.
\cite{Chen3}

The calculated SHG values for the wurtzite GaN and AlN are shown in 
Table \ref{tab:wz}. Again we see that our LMTO results agree quite
well with those obtained in the FLAPW approach\cite{Hughes1}.
An interesting characteristic of the wurtzite material is the 
value of the ratio $\chi_{333}/\chi_{311}$ which is expected 
to be equal to $-2$ in a ``quasicubic'' model. In fact, this
result is obtained if one applies a coordinate axis transformation 
of the cubic tensor from the usual cubic axes to a set of 
rhombohedral axes with the $z$-axis point along the cubic [111] direction. 
Because, there is only one independent component $\chi_{123}^{zb}$ 
in zincblende, this leads to a (geometric)  
relation between the two rhombohedral 
tensor components, which are the same as the hexagonal components.
Based on the similarity of the bonding in wurtzite and zincblende,
this also predicts a relation between 
the values of $\chi^{(2)}$ in wurtzite and the  value in zincblende:
$\chi_{333}^{wz}=2\chi_{123}^{zb}/\sqrt{3}$. 
These relations for example are automatically satisfied in the 
bond charge model \cite{BLevine}
independently of the model-parameters. 
The occurence of non-ideal tetrahedral 
bonding clearly produces a deviation from the geometric ratio.\cite{Fujii}
Nevertheless, it is difficult to understand
the large deviation encountered
for AlN. We find, in fact, that even for ideal AlN, these relations  
are strongly violated. This was also concluded in a 
previous discussion of this issue by Chen et al.\cite{Chen1,Chen2}

For GaN all the calculations done by different methods find 
absolute values of the ratio $|\chi_{333}/\chi_{311}|$
that are definitely smaller than two, but an accurate 
experiment of Miragliotta et al. \cite{Miragl}
nevertheless gives a value 
for the ratio very close to $-2$. 
The sign of the SHG components
in this experiment is opposite to all the calculations, probably
because the overall sign was not determined experimentally relative
to atomic coordinates. 
For AlN both theory end experiment give very small
values for $\chi_{311}^{wz}$ and a $\chi_{123}^{zb}$.
The direct comparison of the value of $\chi_{333}$ with experiment
is difficult because the error in the only available experiment, that
of Fujii et al, \cite{Fujii} was very high.
The reasons for the very small components of $\chi^{zb}_{123}$
and $\chi^{wz}_{311}$ in the static limit can be traced back 
to an almost perfect cancellation of inter and intraband contributions
but the reason for this is still not fully understood. 
The question about the ratio of different components in the
hexagonal materials  will be the subject 
of a further detailed study addressing this question for the SiC 
polytypes.\cite{tobepub}

\subsection{Frequency dependent results}

Figure \ref{fig:GaAs_Hughes} shows the comparison of our
calculated values for frequency dependent SHG in
GaAs and GaP with those computed by the FLAPW method
\cite{Hughes}. We used the ``scissors'' approach here;
then, since all aspects of the two calculations are the same except
for the computational
methods, the comparison provides a mutual check of the methods.
The overall agreement in both the positions of all the features
as well as in their amplitudes is very good
over the full 6 eV range considered, especially for GaP.
We note that both ours and the FLAPW results 
differ substantially from earlier work.\cite{Huang1,Moss}
As we mentioned before, in GaAs the LDA gap obtained in LMTO was very 
small, and the gap correction was almost the only source
of the final gap. This could be the reason for differences of SHG
in different LDA approaches because the LDA value of
the gap can be different in slightly different 
implementations of the LDA calculations.
For GaP the LDA gap was substantially larger, and the agreement
between the results is noticeably better.
This agreement provides an additional support
for the accuracy of our approach and 
confirms that different properly implemented band structure methods must
give rather similar results.

However, when one compares the frequency dependent SHG obtained
with the ``scissors'' corrections with those calculated
within the ``shifted'' Hamiltonian approach (see Fig.\ref{fig:GaAs_Ham}),
important differences become apparent. 
The values of the minimal energy at which the resonant transitions
start are obviously the same as the gap was corrected to its
experimental value. But the energies of the main features
in the two curves are obviously different. One can especially
observe that in the energy region between 3 and 6 eV.
The ``scissors'' method obviously rigidly shifts all the conduction bands
to  higher energies.
On the other hand, the shifts in the Hamiltonian only affect
the states with a large s- component so that
the lowest conduction bands are shifted but
the higher bands remain nearly unchanged from their LDA positions. This means
that the states near the gap are probably reproduced better in
the second approach than in ``scissors'' (because of the consistency
of the eigenvectors and eigenvalues) but that the higher bands 
may be incorrect. In fact, we found in several recent studies
of linear optics that the next few higher bands are equally shifted
up.\cite{uvsic,uvgan}
However, as we showed in the last subsection the
shifted Hamiltonian  approach gives
much better results for the zero frequency limit of SHG than the
``scissors'' corrections. This probably reflects the fact that
the low-lying conduction bands are more important for obtaining
the correct value of $\chi^{(2)}$ than 
the whole remaining energy band spectra.
And, the method with ``shifted'' Hamiltonian describes the
matrix elements of transitions into the lowest conduction bands
better than the ``scissors'' approach because of the use of the better
wavefunctions, those which are the eigenfunctions of the corrected
Hamiltonian. 

Figure \ref{fig:GaAs_SHG} shows the calculated frequency dependent
absolute values of $\chi^{(2)}(-2\omega,\omega,\omega)$ for GaAs obtained
on the basis of the ``shifted'' Hamiltonian gap corrections.
The agreement between these calculations and available experimental
results is 
satisfactory considering the nature of the data. 
It is much better than between the ``scissors''
calculation of Ref.\onlinecite{Hughes} and the same collection of
the experimental results. Unfortunately, the experimental
results exist only in the energy
region between 1 and 3 eV (probably due to the available laser light sources)
and there are nonnegligible quantitative differences between
the different sets of data.
Only the electronic transitions into the lowest conduction bands
contribute strongly in the energy range containing the data.
Therefore, our result confirms the fact that the correct values
of the momentum matrix elements into the low lying conduction
bands are important for both the zero frequency values of the
SHG as well as for its low energy (visible light region) values
for GaAs and GaP.
The question about the higher energy region remains open.
To our knowledge there are no measurements of SHG for GaAs
in the higher energy region. This further supports our 
focusing on the low energy region.

Consistent with our results for non-linear response, we find that 
the imaginary part of the
linear dielectric function (DF), $\varepsilon_2(\omega)$, calculated in
the same approach, i.e. with the ``shifted'' Hamiltonian diagonal matrix
elements, agrees excellently with the experiment, as is 
shown in Fig.\ref{fig:GaAs_DF}.
It is also dominated by transitions to the lowest conduction bands.
Therefore, for GaAs, the answer to the question of whether 
the rigid shift of the conduction
band provided by the ``scissors'' corrections is correct in the
higher energy region is not clear. An accurate calculation
of the self-energy corrections to the LDA bands can help us to
resolve the question. Further complications for these higher
bands may in fact arise from local field and continuum excitonic 
effects both of which may shift oscillator strength. 
This is actually also true for the lower bands and thus our 
good agreement  with experiment  may to some extent be due to 
effective cancellations of effects neglected in the present treatment.
In the remainder of this paper, we use the simpler scissor's approach
either to facilitate comparison to other work or for other reasons
previously mentioned.

The imaginary part of the three independent SHG components (333, 311, and 131) 
for wurtzite GaN and AlN are shown in Figure \ref{fig:GaNAlN}.
The results are very similar to those obtained in Ref.\onlinecite{Hughes1}
as is to be expected since the same ``scissors'' approach is used.
We first note a strong similarity between the 311 and 131 components
over the entire energy spectrum. This is consistent with their exact 
equality for the static value, an expression of the Kleinman symmetry.
Except for the structure at the two-photon threshold,
the imaginary part of the SHG
function is rather flat and small for all the components in a region of
approximately 1 eV above the threshold. In the higher energy
region, features with much larger amplitude appear.
Also, the positions of the features are approximately
the same for all the polarizations. But the signs 
of the 311 and 131 components 
are usually opposite to those for the 333 component. This fact
is consistent with the opposite signs of the zero frequency SHG components
which can be obtained from the frequency dependent imaginary parts
by use of the Kramers-Kronig transformation. 

As mentioned in the previous subsection, the static $\chi_{123}^{zb}$
and $\chi_{311}^{wz}$ components are very small for AlN.
Fig. \ref{fig:aln} shows that this results from an almost
exact cancellation between the interband and intraband contributions
over the entire energy range. 

Next, we turn to our results for the frequency dependent 
SHG in zincblende (or 3C) SiC,
which we believe has not been calculated before.
In SiC, the use of 1 eV rigid shift of all conduction bands
allows one to reproduce the
positions of all the main features of the linear dielectric function
over a very wide energy region (up to 10 eV) \cite{uvsic}
for the most abundant polytypes (3C, 4H, 6H, and 15R).
Therefore, the ``scissors'' corrections seem more appropriate here
than adjusting only the lowest band. Furthermore, it is expected to
work much better for this material than for GaAs in view of the 
much smaller relative gap correction.

Fig.~\ref{fig:SiC_ei} shows the imaginary part of the SHG for 3C-SiC
together with its interband and 
intraband components.
As one can see, these contributions have comparable absolute
values but different signs in the wide energy region. This means that
the reasonable description of the total SHG can not be achieved
if one of these components is ignored.

It is also of interest to separately show and analyze
the $2\omega$ and $\omega$ resonant contributions to the SHG.
The total SHG curve is the sum of these components, both of which have
features at different energies 
thus complicating the interpretation.
The comparisons of the $2\omega$ and $\omega$ contributions to the
$\chi_{123}$ component of the SHG to the imaginary
part of the DF are given in panels a) and b) of Fig. \ref{fig:SiC_w2w},
respectively.
We compare the $2\omega$ term with
$\varepsilon_2(2\omega)$ and the $\omega$ term with $\varepsilon_2(\omega)$
with the DF's suitably multiplied by scaling factors.
First, we note that the energies of all the main structures and 
peaks in the SHG and DF curves coincide 
because the electronic transitions
which produce the structures in both the curves are the same, only
the matrix element factors in the two quantities are
different. The origin of the peaks can be analyzed by decompositions
of $\varepsilon_2$ and $\chi^{(2)}$ into separate band-to-band
contributions
(see, e.g., Ref.\onlinecite{uvsic} for details of such an analysis of
the linear DF). 
This exercise illustrates the utility of decomposing the total $\chi^{(2)}$
into its $\omega$ and $2\omega$ resonance components 
each of which is closely
related to the linear optical spectrum.
Because the linear response function is more readily 
understood in terms of specific band-to-band components, the decomposition
is useful for extracting 
the band structure information from $\chi^{(2)}$
spectra. We will pursue this approach in more detail in a future 
publication.\cite{tobepub}
We merely point out here that the additional fine 
structure visible in ${\rm Im}\chi^{(2)}$ and 
resulting from the sign variations 
in the matrix elements may be helpful in identifying specific band-to-band
features. In a sense, the additonal power of the electric field in the second 
harmonic generation produces a modulation. The resulting interference 
effects lead to more structured spectra in a manner similar to 
electroreflectance or other modulation spectroscopies.

\section{Conclusions} \label{sec:con}

In summary, we have shown in this paper that the LMTO method in the ASA
allows us to perform calculations of non-linear response functions 
with the accuracy of that obtainable by other first-principles methods.
At present, we have implemented the approach for second harmonic generation,
including its full frequency dependence in the absorbing region within
the one electron approximation in the long-wave length limit, i.e. 
without inclusion of local field effects. This limitation can be removed
in future work. The present implementation was carried out using the
equations derived by Aversa and Sipe\cite{Aversa} 
in the length-gauge formalism, 
or, equivalently, using a separation of 
strictly interband and mixed inter- and intraband processes 
as derived by Sipe and Ghahramani\cite{Sipe}. This has the advantage
that these contributions can be studied separately. Furthermore,
with a relatively simple rearrangement of the terms in the static limit
the equations explicitly exhibit the Kleinman symmetry.
These equations are applicable to crystals of any symmetry
unlike some of the earlier formulations which were restricted to 
crystals of cubic symmetry. 

We found excellent agreement with other ab-initio results when care is taken 
to use the same approximations for the underlying electronic structure.
This was shown most convincingly for GaAs, GaP, and wurtzite GaN and AlN 
where it was possible to compare with calculations \cite{Hughes,Hughes1} 
which differed 
from ours solely in the band structure method employed.  Excellent agreement
was obtained for these materials both for the static values 
and for the frequency dependence.  
Our static values are also in good accord with those obtained by an
approach 
introduced by Levine and Allan \cite{Levine1,Levine2,Levine3,Chen1,Chen2}
which differs significantly from ours in the details of its implementation.

On the other hand, we find the results to be quite sensitive to  
the particular manner in which band gap corrections are applied.
We found that the commonly used ``scissors'' approach
has some deficiencies even when using the associated renormalization 
of the momentum matrix elements. We identified the loss of consistency between 
eigenfunctions and eigenvalues in this approach as a reason for 
underestimating the momentum matrix elements. We showed that a somewhat
different approach, in which corrections are made to the one electron 
Hamiltonian matrix which is then explicitly rediagonalized, improves 
the agreement with experiment both in the static limit and in the 
low frequency absorbing region. At present, the latter approach 
is applied essentially only to the lowest conduction bands. As such, it is 
not expected to improve the LDA for the features relating to higher
energy bands. Because these are found generally to be of less importance 
for the optical spectra in the low energy region accessible to 
current experimental techniques (even more strongly so in SHG 
than in linear optics), this is not considered to be 
an important practical limitation. 

It is worth noting that this illustrates another advantage of the LMTO method.
Its use of a physically transparent atomic orbital like basis set 
is what allowed us to implement rather easily some ad-hoc corrections 
to the Hamiltonian which improve the band structure and capture 
the essential aspects of the physics. Similarly, 
other corrections to the LDA 
such as those required for highly correlated d-band and f-band 
systems in what is known as the LDA$+$U approach
are relatively easily included in LMTO based methods.  
While our present approach for including corrections to the LDA for 
semiconductors is not perfect and rather ad-hoc, there is hope that 
more rigourous approaches such as those based on the GW approximation 
\cite{Aryasetiawan} or on the screened exchange approximation 
\cite{Rucker,Maksimov}   can be implemented in a similar manner.

In the process of testing our approach, we also found that the 
effective mass sum rule is satisfied typically to within 10 \%
by our LMTO-ASA computed momentum matrix elements and eigenvalues.
This generally required including a somewhat higher angular momentum 
cut-off than that used in usual electronic structure calculations.
The results for $\chi^{(2)}$ were generally found to converge even 
better than these sum rules showing that the latter 
constitutes a rather strict test of the accuracy of the momentum 
matrix elements.  
For the calculation of static values of  $\chi^{(2)}$, 
we derived a separate set of equations which contain less singular 
terms to be summed over the Brillouin zone. By 
employing a  semi-analytical method in combination with 
linear interpolation of the energies and the matrix elements
inside each microtetrahedron, an efficient integration scheme 
over the Brillouin zone was obtained. This scheme requires fewer 
${\bf k}$ points than that used 
for the frequency dependent response functions.

The convergence of the SHG coefficients 
with respect to  the number of bands included in the summation
was also found to be very good, and shows that the results are 
dominated by  the low-lying bands. This reduces concerns about 
the accuracy of the higher lying bands in linear methods. 

Decomposing ${\rm Im}\chi^{(2)}(-2\omega,\omega,\omega)$ 
into $\omega$ and $2\omega$ resonant components 
was shown to be helpful in relating the band structure to the SHG spectra. 
In was found that the method reported here
for calculating $\chi^{(2)}$ based on the LMTO codes is fast, flexible,
and accurate.
This approach should allow us to study the non-linear optical properties
of systems containing a large number of atoms 
per unit cell such as heterostructures and superlattices.

\acknowledgements
We thank Dr. James Hughes and Prof. John Sipe 
for sending their manuscript
on GaN and AlN to us prior to publication. 
Part of the computations were performed at the Ohio Supercomputer Center.
This work was supported in part by  NSF (DMR-95-29376) and 
AFOSR (F49620-95-1-0043).

%
%

\begin{table}
\caption{ Calculated longitudinal effective masses, $m^{*}/m$, for a few
valence and conduction bands of 3C-SiC at the X-point of the BZ. 
MME - the masses given by Eq.(\ref{eq:sum2}) using
calculated the ASA momentum
matrix elements and band energies,
for different $l_{max}$, the maximum angular momentum included in the LMTO
Hamiltonian. The last column gives the values of $m^{*}/m$ calculated
directly by differentiating of the band energies $E(\bf{k})$ obtained in
the ASA scheme. }
\label{tab:mass}
\begin{tabular}{lcccc}
Band      &  MME($l_{max}$=2) & MME($l_{max}$=3)  &  MME($l_{max}$=4) & 
           $E^{ASA}(\bf{k})$  \\\tableline
$X_1^v$ & -0.185 & -0.181 & -0.183 & -0.155 \\

$X_3^v$ &  0.141 &  0.146 &  0.148 &  0.138 \\

$X_1^c$ &  0.579 &  0.557 &  0.576 &  0.619 \\

$X_3^c$ &  0.945 &  0.951 &  0.938 &  0.928 \\
\end{tabular}
\end{table}

\begin{table}
\caption{LDA calculated values of the static $\chi^{(2)}$ for 3C-SiC 
for different $l_{max}$ and $n_{max}$ (in pm/V). All the results were
obtained using 735 {\bf k} points in an irreducible wedge of the BZ.}
\label{tab:conv}
\begin{tabular}{lccccc}
$l_{max} \backslash n_{max}$ & 10 & 15 & 20 & 25 & 30 \\
\tableline
2 (spd)   & 9.12  & 11.53 & 11.80 & 11.92 & 11.96 \\
3 (spdf)  & 15.28 & 17.31 & 17.68 & 17.79 & 17.82 \\
4 (spdfg) & 15.09 & 17.25 & 17.49 & 17.60 & 17.65 \\
\end{tabular}
\end{table}

\begin{table}
\caption{Calculated values of the static $\chi^{(2)}$ for the zincblende 
semiconductors GaAs and GaP (in pm/V). 
The results of the present calculations -
LDA, LDA+``scissors'', LDA + shift of corresponding Hamiltonian
matrix elements (LDA+$\Delta H$) -
are compared with other theoretical results and experimental data. 
LF stands for the local field effects,
NCC - for non-linear core-correction. The experimental values 
are rescaled as discussed in Ref. \protect\onlinecite{Levine3}.
}
\label{tab:GaAs}
\begin{tabular}{llcc}
 & & & \\
	 &  Method    &    GaAs    &   GaP    \\
\hline
Present  &  LDA                & 735.6  &   153.4  \\
	 &  LDA + ``scissors'' & 104.8  &  48.0  \\
	 &  LDA + $\Delta H$   & 162.0  &  65.1  \\
 & & & \\
Levine \protect\cite{Levine3} &  LDA+LF, no ``scissors''
			      & 354 & 129.2 \\
			  &  LDA+``scissors'', no LF & 184.6 & 84.2 \\
			  &  LDA+``scissors'' + LF & 172.4 & 75 \\
 & & & \\
Dal Corso et al \protect\cite{DalCorso} & LDA+LF, NCC, no ``scissors'' 
					& 205    &  83 \\
                                        & LDA+LF, no NCC, no ``scissors''
					& 158 & 68 \\
 & & & \\
Hughes and Sipe \protect\cite{Hughes} & LDA+``scissors'', no LF 
				      &  96.5   &  50.3 \\
 & & & \\
Huang and Ching \protect\cite{Huang1} & LDA    & 251  & 135 \\
 & & & \\
Levine and Bethea \protect\cite{Levine} &  Experiment at $\hbar\omega=0.117$ eV 
				      & 162 $\pm$ 10 & 74 $\pm$ 4 \\ 
\end{tabular}
\end{table}

\begin{table}
\caption{
LDA and LDA+ ``scissors'' calculations of the static $\chi^{(2)}$ for
several wide band gap semiconductors with the zincblende structure
compared with pseudopotential
LDA calculations which included the local field (LF) effect (in pm/V).
}
\label{tab:zb}
\begin{tabular}{lccccc}
 & & & & & \\
   &  Method  &    GaN   &  AlN  &  BN  &  3C-SiC  \\
\hline
Present &  LDA              &  18.3 & -0.4  &  4.5  &  17.6 \\
	&  LDA+``scissors'' &  10.6 & -0.23 &  2.4  &  11.3 \\ 
 & & & & & \\
Chen et al \protect\cite{Chen1,Chen2} & LDA+LF 
			    &  16.9 & 0.02  &  5.6  & 24.4  \\
\end{tabular}
\end{table}

\begin{table}
\caption{ 
Calculated values of the static $\chi^{(2)}$ 
for wurtzite wide band gap semiconductors GaN and AlN (in pm/V)
compared with other calculations and experimental results. The values
of the 333
and 311 SHG components as well as their ratios are shown.
}
\label{tab:wz}
\begin{tabular}{ll|ccc|ccc}
 & & & & & & & \\
   &  Method  &     &  GaN  &   &   & AlN  &    \\
\tableline
 & & & & & & & \\
   &   &  $\chi_{333}$ & $\chi_{311}$ & $\chi_{333}/\chi_{311}$
       &  $\chi_{333}$ & $\chi_{311}$ & $|\chi_{333}/\chi_{311}|$   \\
\hline
Present &  LDA              & 8.8 & -7.1 & -1.24 & -7.5  & +0.4  & 18.75 \\
	&  LDA+``scissors'' & 5.6 & -4.3 & -1.3  & -4.3  & +0.24 & 17.9  \\ 
 & & & & & & & \\
Hughes and Sipe \protect\cite{Hughes1} & LDA+``scissors'' 
			    & 6.03 & -4.27 & -1.41 & -3.77 & -0.25 & 15.1 \\
 & & & & & & & \\
Chen et al \protect\cite{Chen2} & LDA+LF, no ``scissors''
			    & 11.52 & -6.9 & -1.67 & -9.22 & -0.48 & 19.2 \\
 & & & & & & & \\
Miragliotto et al \protect\cite{Miragl} & Experiment 
			    & -10.7 & 5.3  & -2.02 &  -   &  -   & - \\
 & & & & & & & \\
Fujii et al \protect\cite{Fujii}  & Experiment 
	   &   -    &  -  & - & -12.6$\pm$7 & ~$\leq |$ 0.5 $|$ & - \\
\end{tabular}
\end{table}

\begin{figure}
\begin{center}
\caption{
The calculated ${\rm Im}\chi^{(2)}_{123}(-2\omega,\omega,\omega)$
for GaAs and GaP (solid lines) compared with the
FLAPW results of Ref.\protect\onlinecite{Hughes} (dotted lines).
Both the LMTO and FLAPW approaches included the ``scissors''
corrections to the LDA gap.
}
\label{fig:GaAs_Hughes}
\end{center}
\end{figure}

\begin{figure}
\begin{center}
\caption{Comparison of the ${\rm Im}\chi^{(2)}_{123}(-2\omega,\omega,\omega)$
for GaAs and GaP calculated within different approaches
to the gap corrections in LDA: the shift of corresponding
diagonal Hamiltonian matrix elements 
(solid lines, see the text); 
the ``scissors'' corrections (dotted lines).
}
\label{fig:GaAs_Ham}
\end{center}
\end{figure}

\begin{figure}
\begin{center}
\caption{Calculated $|\chi^{(2)}_{123}(-2\omega,\omega,\omega)|$
for GaAs (solid line) compared with experiments 
of Ref.\protect\onlinecite{Parsons}
(dotted line), Ref.\protect\onlinecite{Bethune} (dashes), 
and Ref.\protect\onlinecite{Chang}
(filled circles).}
\label{fig:GaAs_SHG}
\end{center}
\end{figure}

\begin{figure}
\begin{center}
\caption{Calculated 
$\varepsilon_2(\omega)$ for GaAs (solid line) compared with 
experiment \protect\cite{Philipp} (dotted line).}
\label{fig:GaAs_DF}
\end{center}
\end{figure}

\begin{figure}
\begin{center}
\caption{Calculated ${\rm Im}\chi^{(2)}(-2\omega,\omega,\omega)$
for GaN and AlN in the wurtzite structure:
the 333 component (solid line), 311 component (dotted line), 
131 component (dashed line).
}
\label{fig:GaNAlN}
\end{center}
\end{figure}

\begin{figure}
\begin{center}
\caption{
Decomposition of ${\rm Im}\chi^{(2)}_{123}(-2\omega,\omega,\omega)$
for AlN in the zincblende structure in its inter- and intraband components:
the total SHG (solid line), the interband contribution (dotted line),
and the intraband contribution (dashed line).
}
\label{fig:aln}
\end{center}
\end{figure}

\begin{figure}
\begin{center}
\caption{Calculated  
${\rm Im}\chi^{(2)}_{123}(-2\omega,\omega,\omega)$ for 3C-SiC:
the total SHG (solid line), the interband contribution (dotted line),
and the intraband contribution (dashed line).
}
\label{fig:SiC_ei}
\end{center}
\end{figure}

\begin{figure}
\begin{center}
\caption{The lineshapes of the $2\omega$- and $\omega$- resonant contributions
to ${\rm Im}\chi^{(2)}_{123}(-2\omega,\omega,\omega)$
for 3C-SiC compared with the imaginary part of the DF:
a) $2\omega$- contribution to SHG (solid line), in units 10$^{-6}$ esu,
together with $\varepsilon_2(2\omega)/50$ (dotted line);
b) $\omega$- contribution to SHG (solid line), in units 10$^{-6}$ esu,
together with $\varepsilon_2(\omega)/100$ (dotted line).
}
\label{fig:SiC_w2w}
\end{center}
\end{figure}

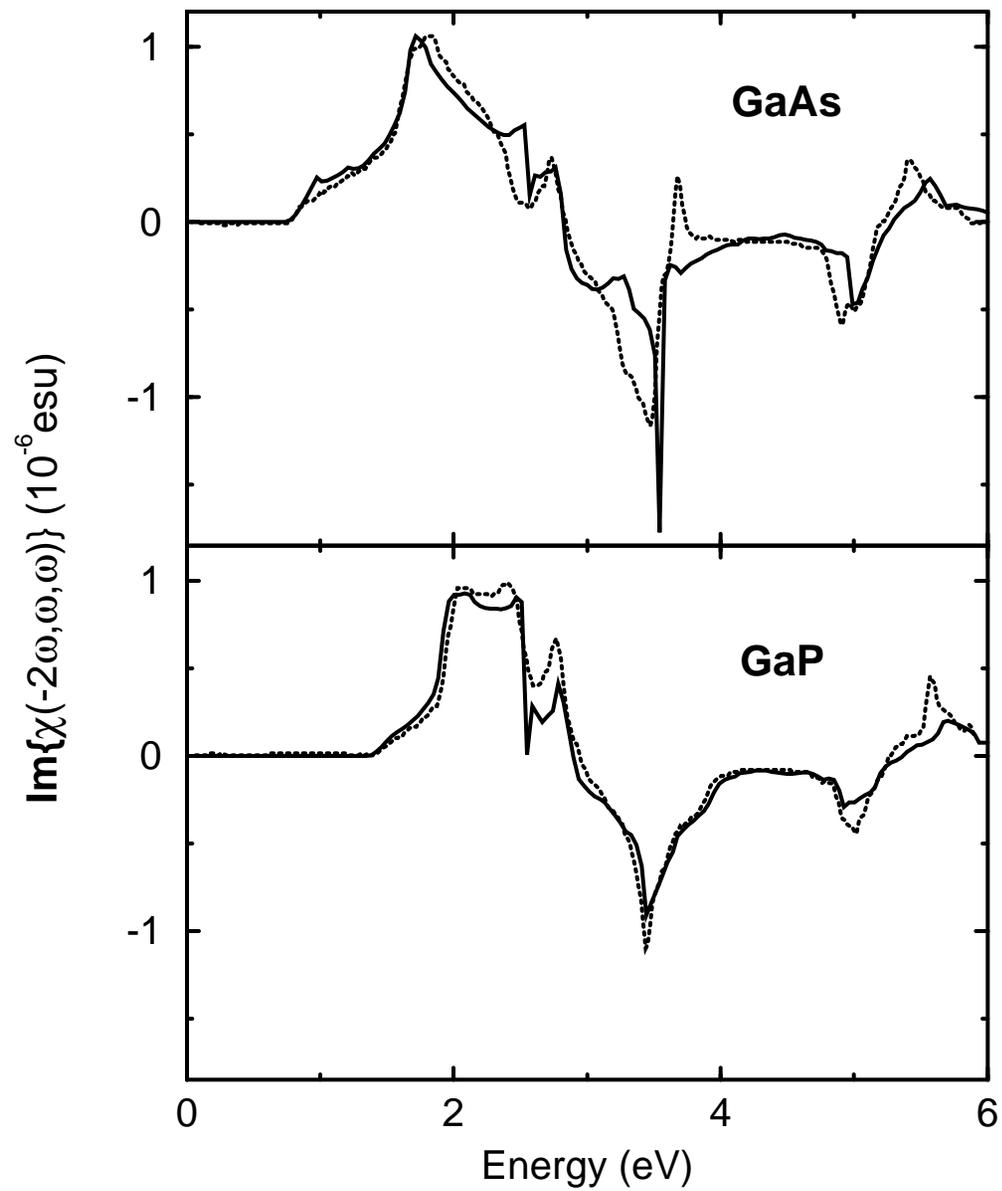
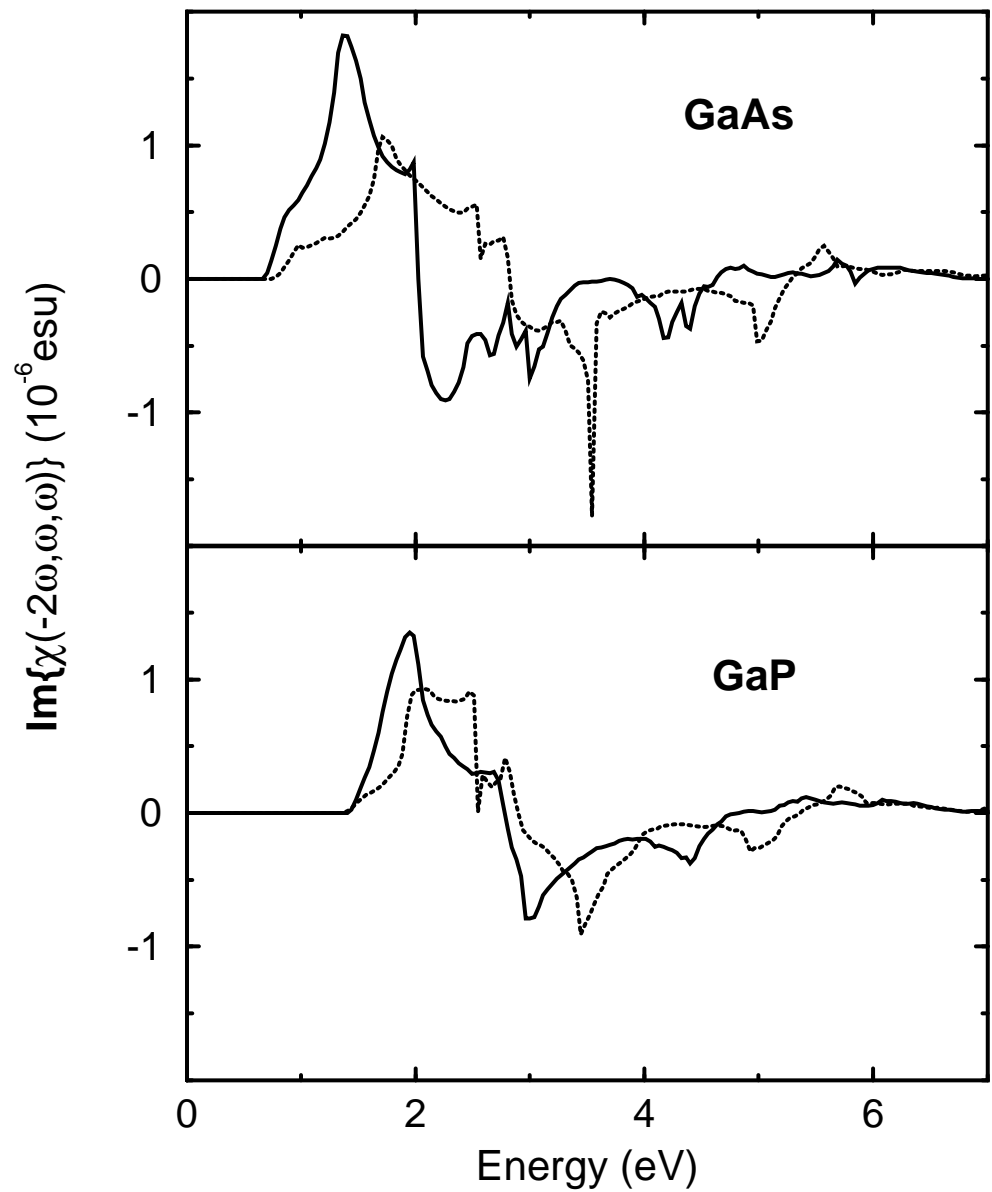
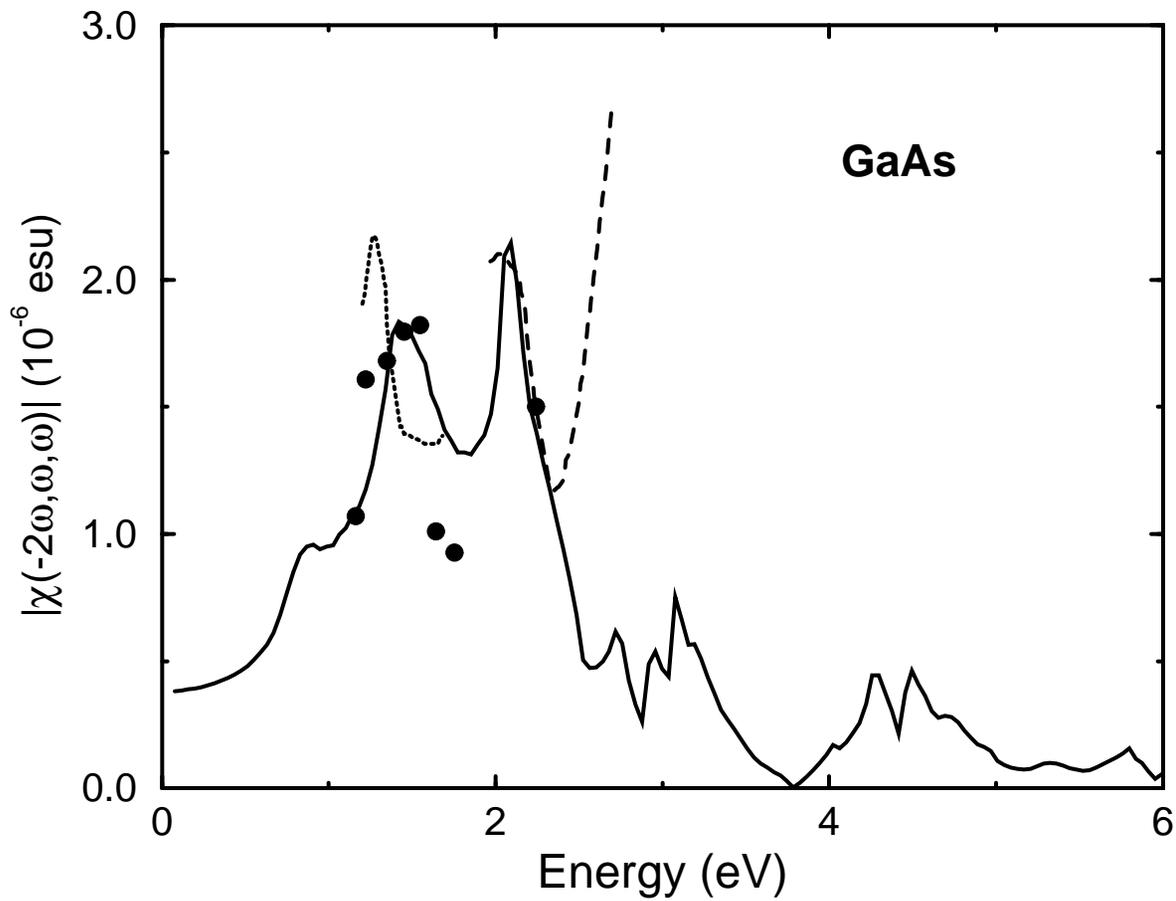
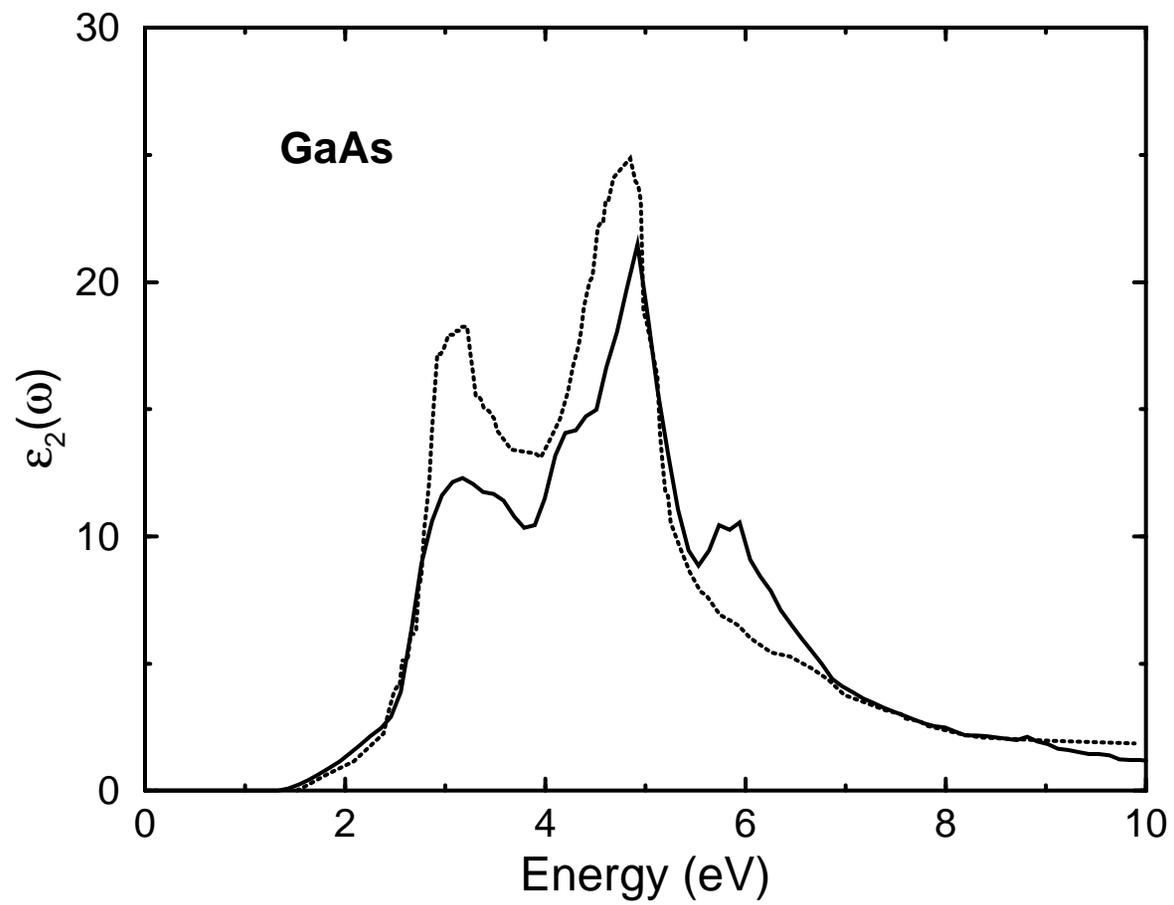
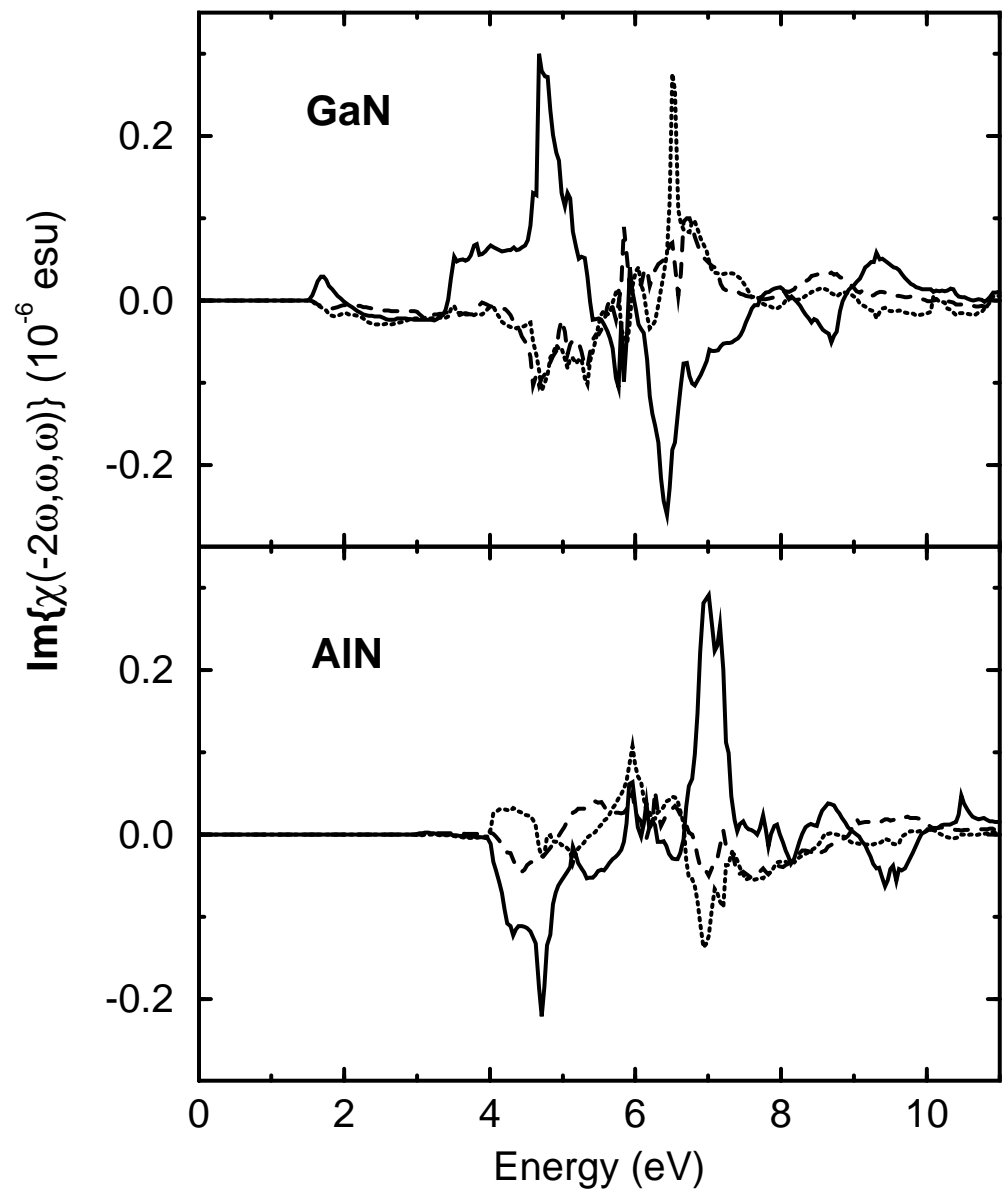
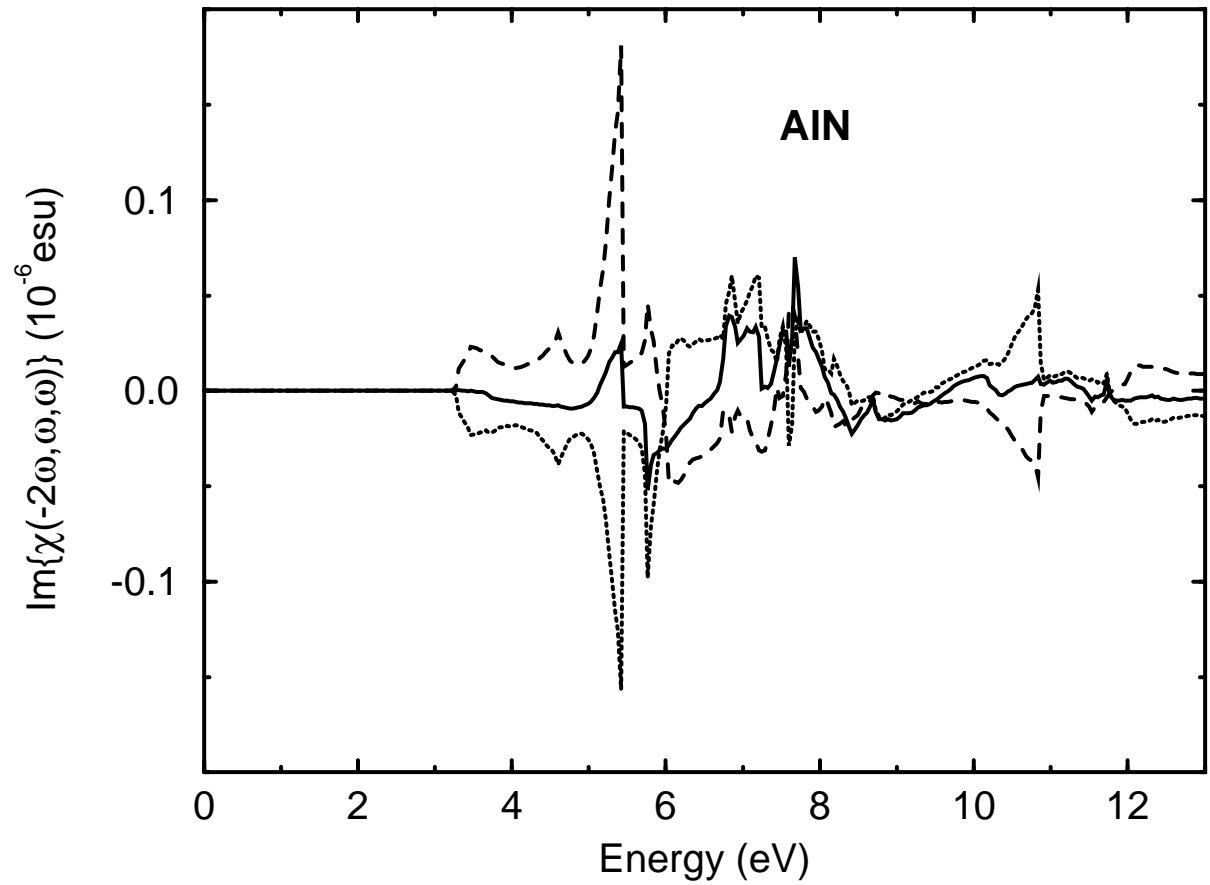
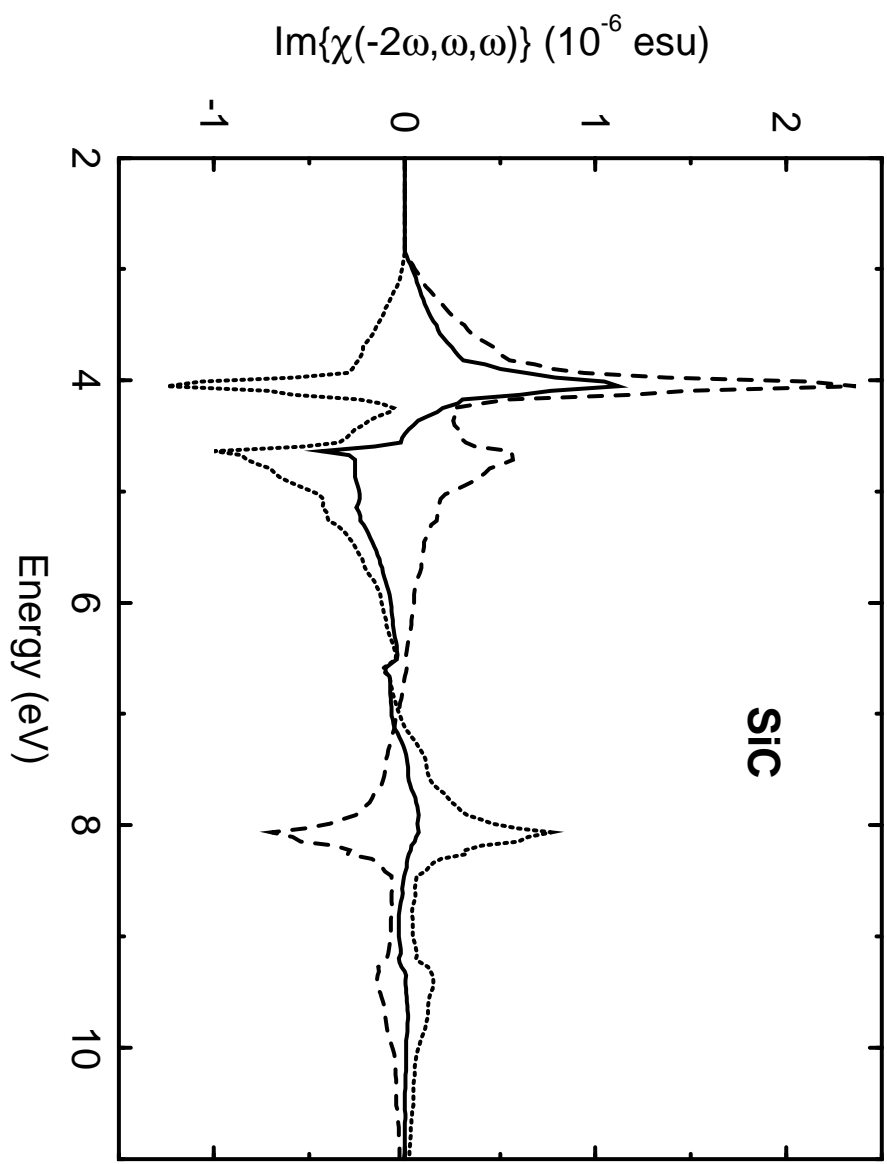
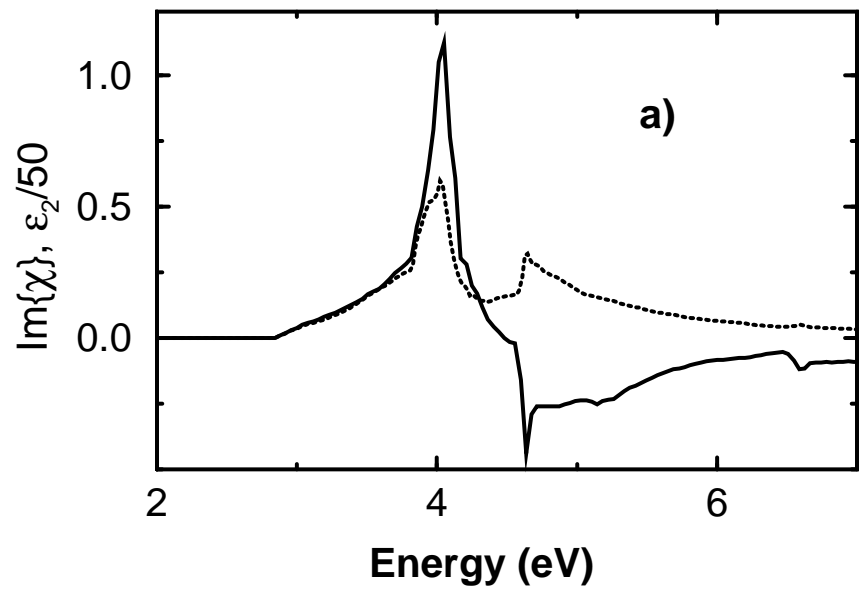
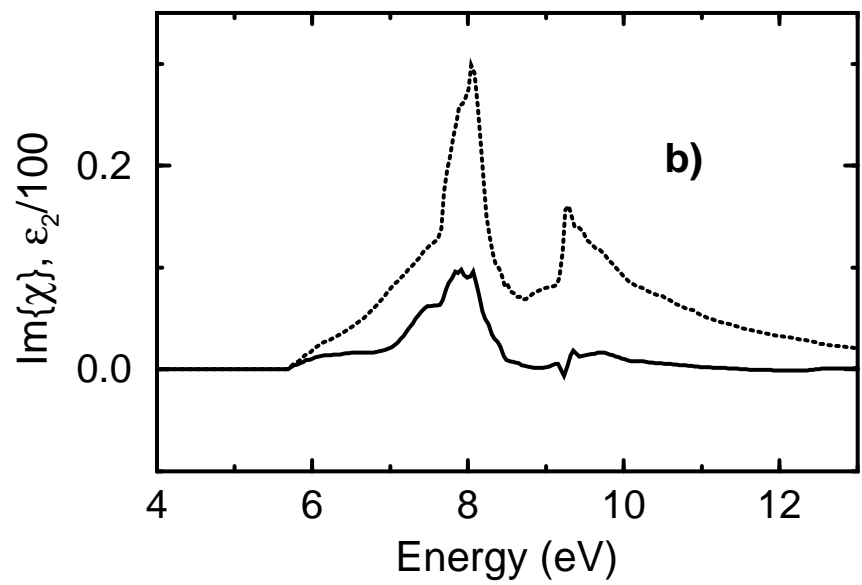
\end{document}